\documentclass[11pt]{article}

\usepackage{a4wide}
\usepackage{graphicx}
\usepackage{amsmath}
\usepackage{amssymb}
\usepackage{hyperref}

\newcommand{\order}[1]{{\cal O}\left(#1\right)}
\newcommand{\comment}[1]{[\textbf{#1}]}
\newcommand{\as}{\alpha_s}
\newcommand{\cone}{\text{SISCone}}
\newcommand{\cam}{\text{Cam}}
\newcommand{\kt}{k_t}
\newcommand{\cNg}{{\cal N}_g}
\newcommand{\cV}{{\cal V}}
\newcommand{\ie}{{\it i.e.}\ }
\newcommand{\cf}{{\it cf.}\ }
\newcommand{\eg}{{\it e.g.}\ }
\newcommand{\GeV}{\,\mathrm{GeV}}
\newcommand{\TeV}{\,\mathrm{TeV}}
\newcommand{\mb}{\,\mathrm{mb}}
\newcommand{\JA}{\text{JA}}
\newcommand{\bs}[1]{\boldsymbol{#1}}
\newcommand{\gghosts}{\,|\,\{g_i\}}

\newcommand{\avg}[1]{\left\langle #1 \right\rangle}

\newcommand{\PU}{\mathrm{PU}}

\newcommand{\oPJ}{\mathrm{1PJ}}

\title{\textbf{The Catchment Area of Jets}}

\author{Matteo Cacciari, Gavin P.~Salam \\
  \small \it LPTHE\\
  \it\small UPMC Universit\'e  Paris 6,\\
  \it\small Universit\'e Paris Diderot -- Paris 7,\\ 
  \it\small CNRS UMR 7589, Paris, France\\[10pt]
  Gregory Soyez\\
  {\it\small Brookhaven National Laboratory, Upton, NY 11973, USA}\\[5pt]
}

\date{}

\begin{document}

\maketitle
\vspace{-8.5cm}
\begin{flushright}
  February 2008\\
  LPTHE-07-02\\
  arXiv:0802.1188
\end{flushright}
\vspace{6.5cm}

\begin{abstract}
  The area of a jet is a measure of its susceptibility to radiation,
  like pileup or underlying event (UE), that on average, in the jet's
  neighbourhood, is uniform in rapidity and azimuth.
  In this article we establish a theoretical grounding for the
  discussion of jet areas, introducing two main definitions, passive
  and active areas, which respectively characterise the sensitivity to
  pointlike or diffuse pileup and UE radiation.
  We investigate the properties of jet areas for three standard jet
  algorithms, $k_t$, Cambridge/Aachen and SISCone. Passive areas
  for single-particle jets are equal to the naive geometrical
  expectation $\pi R^2$, but acquire an anomalous dimension at higher
  orders in the coupling, calculated here at leading order. The more
  physically relevant active areas differ from $\pi R^2$ even for
  single-particle jets, substantially so in the case of the cone
  algorithms like SISCone with a Tevatron Run-II split--merge procedure.
  We compare our results with direct measures of areas in
  parton-shower Monte Carlo simulations and find good agreement with
  the main features of the analytical predictions. 
  We furthermore justify the use of jet areas to subtract the contamination from
  pileup.
\end{abstract}

\newpage
\tableofcontents

\section{Introduction}
\label{sec:introduction}

For nearly three decades now, jets \cite{SW77} have represented the
principal tool for accessing information about an event's partonic
hard-scattering structure and kinematics. As a result of this, much
work has been carried out on understanding the properties of jets in a
range of collider contexts, addressing issues such as jet substructure
\cite{eeResum,hhJetShapes,PerezRamos:2007cr}, the correlations between multi-jet
production and the hard colour-structures present in an event
\cite{ApplebySeymour,BanfiDasgupta}, and perturbative threshold
corrections to jet production~\cite{Threshold}.

One issue that has been largely neglected, but that is highly relevant
in a hadron collider context such as Tevatron or LHC, is that of the
modification of jet kinematics by non-perturbative effects associated
with the proton beams.
These effects, often referred to as a whole as the ``underlying
event'' (UE), are rather poorly understood. However, from tuned 
underlying event models \cite{tuneA,Jimmy} one consistently finds that
a principal effect of the underlying event is to add a rather large
amount of transverse momentum, fairly uniformly throughout the event
--- according to models $3-5\GeV$ per unit rapidity at the Tevatron,
$10-15\GeV$ per unit rapidity at LHC (see \eg \cite{CDMS}), with substantial fluctuations in
the amount from one event to another. This is an order of magnitude
larger than the normal scale for non-perturbative effects in an
$e^+e^-$ or DIS context, which amount to $\sim 0.5\GeV$ per unit
rapidity (with respect to the $q\bar q$ axis)
\cite{BryanTube,DSReview}. Thus for jets with transverse momenta,
$p_t$, of several tens of $\GeV$, the UE can be as significant as the
perturbative corrections to the jet transverse momentum $\as p_t$.
A related issue is that of pileup (PU), minimum-bias collisions that
occur in the same bunch crossing as the main event, with the potential of
further adding up to $100 \GeV$ of soft radiation per unit rapidity.

Given the large momentum scales associated with the underlying event
and pileup, it is important to develop tools for understanding how
they both affect jets. Of the two, PU is conceptually simpler because
the particles it adds are entirely uncorrelated with those of the hard
scatter. In contrast UE particles cannot entirely be disentangled from
those due to the hard scatter (neither theoretically, nor in
practice). Nevertheless it is useful, and probably not too poor an
approximation, to treat the UE as largely independent from the hard
scatter, just like PU.

The UE and PU can affect a jet in two ways. Firstly, particles from the
UE and PU may be clustered into the jet, increasing its transverse
momentum. Secondly, the presence of the UE/PU particles can modify the
way in which the particles from the hard scatter get clustered into
jets.

To study the first of these effects, we shall introduce the concept of
the `jet area'. The logic that motivates this particular quantity is
that UE/PU particles are distributed uniformly in rapidity and
azimuth, at least after averaging over many events.  Therefore a good
measure of a jet's susceptibility to contamination by the UE and PU
should be given by the extent of the region (on the rapidity-azimuth
cylinder) over which it is liable to capture UE and pileup particles,
\ie the jet's ``catchment area''.

One might naively think of this area as the area of the
surface covering all the particles that make up the jet. However, a
moment's reflection shows that this area is actually zero, the jet
being made of pointlike particles.
An alternative obvious definition would be to take the area of the
convex hull surrounding all the particles in a jet. However this
definition also fails to satisfy basic sensibleness requirements:
for example, if two jets share an irregular boundary it is possible that
their convex hulls will overlap and the assignment of area to one or
other of the jets becomes ambiguous.

Two main definitions of the jet area can be introduced without such
shortcomings. 
A first one, the \emph{passive area}, involves scanning a single
infinitely soft particle (a ``ghost'') over the rapidity-azimuth
cylinder and determining the region over which it is clustered within
a given jet. It can be understood as a measure of the susceptibility
of the jet to contamination from an UE with pointlike structure. It is
discussed in detail in section~\ref{sec:passive}.
A second definition, studied in section~\ref{sec:active}, the
\emph{active area}, involves adding a dense coverage of infinitely
soft ``ghosts'' and counting how many are clustered inside a given
jet.  It can be understood as a measure of the susceptibility of the
jet to contamination from an UE with uniform, diffuse structure.
The passive and active areas differ because in the latter case the
ghosts can cluster among themselves as well as with the actual event
particles, thus playing a more active role in the clustering.
We will find that the numerical value of the jet area can differ
according to the kind of particles which make up the jet. The simplest
case to study will be that of jets with a single hard particle. We
shall then consider how the jet's area changes when its main hard
parton emits an additional soft gluon, which effectively causes the
jet area to acquire an anomalous dimension.
We will also study ``pure ghost jets,'' those exclusively made up of ghost
particles.

In section~\ref{sec:back-reaction} we shall then consider how the
presence of UE/PU affects the clustering into jets of particles from
the hard scatter, an effect that we refer to as the
\emph{back-reaction} of the UE/PU on the jet structure. Again we shall
examine this in two limits: pointlike and perfectly diffuse UE/PU.

As well as making approximations for the structure of the UE/PU it
will also be necessary to make simplifying approximations for the
jet structure. Thus we will work in an energy ordered limit, in which
the perturbative event consists of one very hard particle, plus an
optional additional soft perturbative particle. This will be
equivalent to making a leading (single) logarithmic approximation,
truncated at its first non-trivial order in $\as$. For some purposes
we shall also assume a small jet-radius parameter $R$ (\ie the
small-cone approximation of \cite{de Florian:2007fv}).

Despite these many simplifications, the calculations that we present
will be seen to provide considerable insight into the mechanisms at
play in jet clustering in events with UE/PU. In particular they will
help highlight characteristic analytical structures that are common to
a range of rather different jet algorithms, as well as significant
jet-algorithm-dependent differences in the quantitative impact of
these analytical structures. Of the algorithms that we will consider,
two are based on sequential recombination (the inclusive $k_t$
\cite{kt} and Cambridge/Aachen \cite{cam} algorithms), while the third
is a modern, infrared-safe stable-cone algorithm (SISCone
\cite{siscone}) with a Tevatron Run-II type split--merge procedure to
resolve overlapping stable cones. A brief description of the three jet
algorithms is given in Appendix \ref{sec:app_defs}.

To help reinforce the connection between our analytical results and
the (simulated) real world, we shall close the article in
section~\ref{sec:real-life} with Herwig \cite{Herwig} and Pythia
\cite{Pythia} Monte Carlo studies. 
We shall also give the foundations of the use of the area concept for
performing PU subtractions \cite{AreaSubtraction}.

\section{Passive Area}
\label{sec:passive}
Suppose we have an event composed of a set of particles $\{p_i\}$
which are clustered into a set of jets $\{J_i\}$ by some infrared-safe
jet algorithm.

Imagine then adding to the $\{p_i\}$ a single {\sl infinitely soft}
ghost particle $g$ at rapidity $y$ and azimuth $\phi$, and repeating
the jet-finding.
As long as the jet algorithm is infrared safe,  the set of jets
$\{J_i\}$ is not changed by this operation: their kinematics and hard particle
content will remain the same, the only possible differences being the presence
of $g$ in one of the jets or the appearance of a new jet containing only
$g$.

The passive area of the jet $J$ can then either be defined as a scalar
\begin{equation}
\label{eq:passive-area-def}
a(J) \equiv \int dy\,d\phi\; f(g(y,\phi),J)\qquad\qquad  f(g,J) =
\left\{\begin{array}{cc}
1 & g \in J \\
0 & g \notin J \\
\end{array}
\right. \; ,
\end{equation}
which corresponds to the area of the region where $g$ is clustered
with $J$, or as a 4-vector,
\begin{equation}
a_\mu(J) \equiv \int dy\,d\phi\; f_\mu(g(y,\phi),J)\qquad\qquad  f_\mu(g,J) =
\left\{\begin{array}{cc}
g_\mu/g_t & g \in J \\
0 & g \notin J \\
\end{array}
\right. \; ,
\end{equation}
where $g_t$ is the ghost transverse momentum.
For a jet with a small area $a(J)\ll 1$, the 4-vector area has the
properties that its transverse component satisfies $a_t(J) =
a(J)$, it is roughly massless and it points in the direction of
$J$.  For larger jets, $a(J)\sim 1$, the 4-vector area acquires a
mass and may not point in the same direction as $J$. We shall
restrict our attention here to scalar areas because of their greater
simplicity. Nearly all results for scalar areas carry over to 4-vector
areas, modulo corrections suppressed by powers of the jet radius
(usually accompanied by a small coefficient).%
\footnote{
  The above definitions apply to jet algorithms in which each gluon is
  assigned at most to one jet. For a more general jet algorithm (such
  as the ``Optimal'' jet finder of \cite{OJF} or those which perform
  $3\to2$ recombination like ARCLUS \cite{ARCLUS}), then one may define
  the 4-vector area as 
  \begin{equation}
    \label{eq:passive-4vect-area-gen}
    a_\mu(J) = \lim_{g_t \to 0} \frac{1}{g_t} \int dy\, d\phi\;
    (J_{\mu}(\{p_i\},g(y,\phi)) - J_{\mu}(\{p_i\}))\,,
  \end{equation}
  where $J_{\mu}(\{p_i\},g(y,\phi))$ is the 4-momentum of the jet as
  found on the full set of event particles $\{p_i\}$ plus the ghost,
  while $J_{\mu}(\{p_i\})$ is the jet-momentum as found just on the
  event particles.
}

\subsection{Areas for 1-particle configurations}

Consider this definition in the context of three different
jet algorithms: inclusive $k_t$, inclusive Cambridge/Aachen, and the
seedless infrared-safe cone algorithm, SISCone. The definitions of all
three algorithms are summarised in appendix~\ref{sec:app_defs}.

Each of these jet algorithms contains a parameter $R$ which
essentially controls the reach of the jet algorithm in $y$ and $\phi$.
Given an event made of a single particle $p_1$, the passive area of the jet
$J_1$ containing it is $a(J_1) = \pi R^2$ for all three algorithms.

\subsection{Areas for 2-particle configurations}
\label{sec:passive-area-2particle}

Let us now consider what happens to the passive jet areas in the presence of
additional soft perturbative radiation. We add a particle $p_2$ such
that the transverse momenta are strongly ordered,
\begin{equation}
\label{eq:strord}
p_{t1} \gg p_{t2} \gg \Lambda_{QCD} \gg g_t\; ,
\end{equation}
and  $p_1$ and $p_2$ are separated by a geometrical 
distance  $\Delta_{12}
= (y_1-y_2)^2 + (\phi_1-\phi_2)^2$ in the $y$-$\phi$ plane. Subsequently
we shall  integrate over $\Delta_{12}$ and $p_{t2}$. Note that $g_t$ has
been taken to be infinitesimal compared to all physical scales to ensure
that the presence of the ghost particle does not affect the real jets.

For $\Delta_{12} = 0$ collinear safety ensures that the passive area  
is still equal to $\pi R^2$ for all three algorithms. However, as one increases
$\Delta_{12}$, each algorithm behaves differently.

\subsubsection{$k_t$}

Let us first consider the behaviour of the $k_t$ jet algorithm, a sequential
recombination algorithm, which has
2-particle distance measure $d_{ij} = \min(k_{ti}^2,k_{tj}^2)
\Delta_{ij}^2/R^2$ and beam-particle distance $d_{iB} = k_{ti}^2$~\cite{kt}.
Taking $\Delta_{12} \sim \Delta_{1g} \sim \Delta_{2g} \sim R$ and
exploiting the strong ordering~(\ref{eq:strord}) one has
\begin{equation}
d_{1B} \gg d_{2B} \sim d_{12} \gg d_{g1} \sim d_{g2} \sim d_{gB} \; .
\end{equation}
From this ordering of the $k_t$ distances, one sees that the ghost always
undergoes its clustering before either of the perturbative particles
$p_1$ and $p_2$. Specifically, if at least one of $\Delta_{1g}$ and 
$\Delta_{2g}$ is smaller
than $R$, the ghost clusters with the particle that is geometrically closer.

If both $\Delta_{1g}$ and $\Delta_{2g}$ are greater than $R$ the ghost
clusters with the beam and will not belong to any of the perturbative
jets.

\begin{figure}[th]
\includegraphics[width=\textwidth]{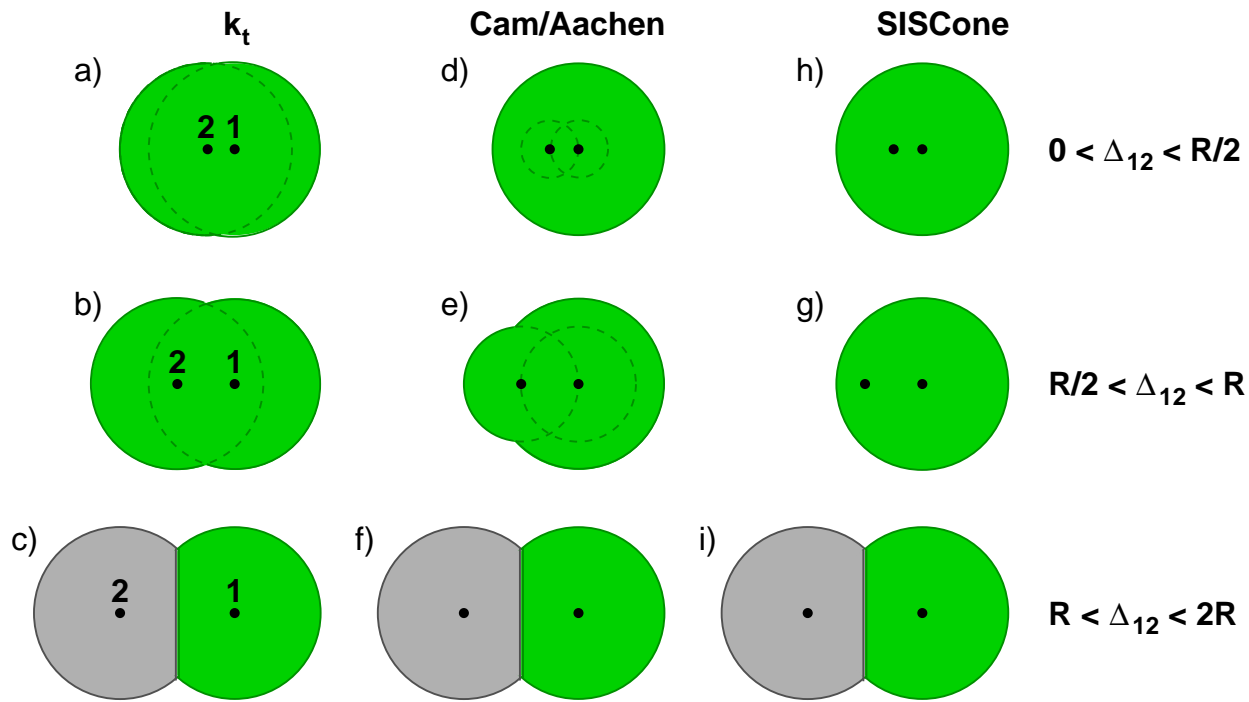}
\caption{Schematic representation of the passive area of a
jet containing one hard particle ``1'' and a softer one ``2'' for various
separations between them and different jet algorithms. Different shadings represent
distinct jets.}
\label{fig:overlaps} 
\end{figure}

Let us now consider various cases. If $\Delta_{12} < R$,
(fig.~\ref{fig:overlaps}a,b)
the particles $p_1$ and $p_2$ will eventually end up in the same
jet. The ghost will therefore belong to the jet irrespectively of
having been clustered first with $p_1$ or $p_2$. The area of the jet
will then be given by union of two circles of radius $R$, one
centred on each of the two perturbative particles,
\begin{equation}
  \label{eq:a_kt_Dlt1}
a_{k_t,R}(\Delta_{12}) =  u\!\left(\frac{\Delta_{12}}{R}\right) \, {\pi R^2}\,,
\qquad\qquad
\mathrm{for}~\Delta_{12} < R\,,
\end{equation}
where
\begin{equation}
\label{eq:u}
u(x) = \frac{1}{\pi}\left[x\sqrt{1-\frac{x^2}{4}} + 2\left(\pi -
\arccos\left(\frac{x}{2}\right)\right)\right],
\end{equation}
represents the area, divided by $\pi$, of the union of two circles of
radius one whose centres are separated by $x$.

The next case we consider is $R < \Delta_{12} < 2R $,
fig.~\ref{fig:overlaps}c. In this case $p_1$ and $p_2$ will never be
able to cluster together. Hence, they form different jets, and the
ghost will belong to the jet of the closer of $p_1$ or $p_2$. The two
jets will each have area
\begin{equation}
  \label{eq:a_kt_Dgt1}
a_{k_t,R}(\Delta_{12}) =   \frac{u\!\left(\Delta_{12}/R\right)}{2}\, \pi R^2\,,
\qquad\qquad
\mathrm{for}~R<\Delta_{12} < 2R\,.
\end{equation}
Finally, for $\Delta_{12} > 2R$ the two jets formed by $p_1$ and $p_2$
each have area $\pi R^2$.

The three cases derived above are summarised in
table~\ref{tab:passive-areas} and illustrated in
fig.~\ref{fig:2point}.

\subsubsection{Cambridge/Aachen}

Now we consider the behaviour of the Cambridge/Aachen jet
algorithm~\cite{cam}, also a sequential recombination algorithm,
defined by the 2-particle distance measure $d_{ij} =
\Delta_{ij}^2/R^2$ and beam-particle distance $d_{iB} = 1$.  Because
in this case the distance measure does not involve the transverse
momentum of the particles, the ghost only clusters
first if $\min(\Delta_{1g},\Delta_{2g}) < \Delta_{12}$. Otherwise,
$p_1$ and $p_2$ cluster first into the 
jet $J$, and then $J$ captures the ghost if $\Delta_{Jg} \simeq
\Delta_{1g} < R$.

The region 
$\Delta_{12} < R$ now needs to be separated into two parts, $\Delta_{12} < R/2$
and $R/2 < \Delta_{12} < R$.

If the ghost clusters first, then it must have been contained in either of the
dashed circles depicted in figs.~\ref{fig:overlaps}d,e. If $\Delta_{12} < R/2$
both these circles are contained in a circle of radius $R$ centred on $p_1$
(fig.~\ref{fig:overlaps}d), and so the jet area is $\pi R^2$.

If $R/2 < \Delta_{12} < R$ (fig.~\ref{fig:overlaps}e)
the circle of radius $\Delta_{12}$ centred on $p_2$ protrudes, and adds to the
area of the final jet, so that
\begin{equation}
{a_{\cam,R}(\Delta_{12})} 
           =  w\!\left(\frac{\Delta_{12}}{R}\right)\, {\pi R^2}\,,
\qquad\qquad
\mathrm{for}~R/2 < \Delta_{12} < R\,,
\end{equation}
where
\begin{equation}
\label{eq:w}
w(x) = \frac{1}{\pi}\left[
\pi-\arccos\left(\frac{1}{2x}\right)+\sqrt{x^2-\frac{1}{4}}+x^2\arccos\left(\frac{1}{2x^2}-1\right)
\right].
\end{equation}
For $\Delta_{12} > R$ a Cambridge/Aachen jet has the same area as
the $k_t$ jet, \cf fig.~\ref{fig:overlaps}f. As with the $k_t$ algorithm, the
above results are summarised in table~\ref{tab:passive-areas} and
fig.~\ref{fig:2point}. The latter in particular illustrates the
significant difference between the $k_t$ and Cambridge/Aachen areas
for $\Delta_{12} \sim R/2$, caused by the different order of
recombinations in the two algorithms.

\subsubsection{SISCone}

Modern cone jet algorithms identify stable cones and then apply a
split/merge procedure to overlapping stable cones. The arguments that
follow are identical for midpoint and seedless cone jet algorithms with a
Tevatron run~II type split--merge procedure~\cite{Blazey}. 
For higher orders, or more realistic events, it is mandatory to use an
infrared-safe seedless variant, a main example of which is
SISCone~\cite{siscone}. 

For $\Delta_{12} < R$ only a single stable cone is found, centred on
$p_1$. Any ghost within distance $R$ of $p_1$ will therefore belong to
this jet, so its area will be $\pi R^2$, \cf fig.~\ref{fig:overlaps}h,g.

For $R <\Delta_{12} < 2R$ two stable cones are found, centred on
$p_1$ and $p_2$. The split/merge procedure will then always split
them, because the fraction of overlapping energy is zero. Any ghost
falling within either of the two cones will be assigned to the closer
of $p_1$ and $p_2$ (see fig.~\ref{fig:overlaps}i). The area of the
hard jet will therefore be the same as for the $k_t$ and Cambridge
algorithms.

\begin{figure}[t]
  \begin{center}
    \includegraphics[width=0.5\textwidth]{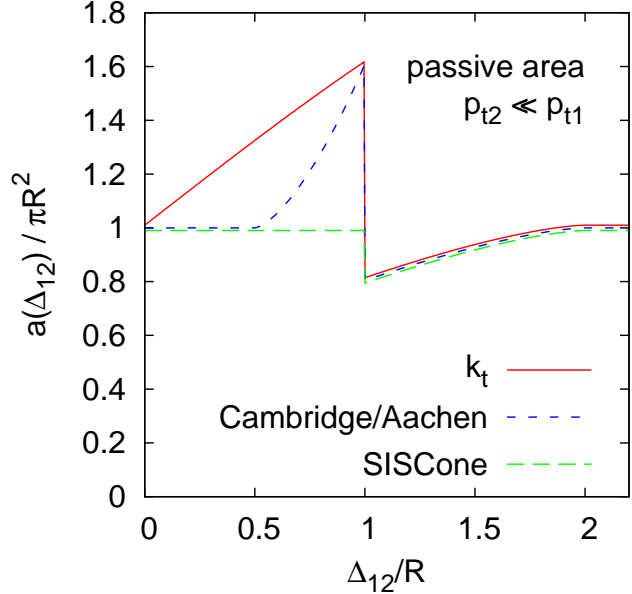}
    \caption{\label{fig:2point} Plot of the passive areas of the hard
      jet as a function of the distance between the hard and the soft
      particle, as given in table \protect\ref{tab:passive-areas}. }
  \end{center}
\end{figure}

Again these results are summarised in table~\ref{tab:passive-areas}
and fig.~\ref{fig:2point} and one notices the large differences
relative to the other algorithms at $R\lesssim 1$ and the striking
feature that the cone algorithm only ever has negative corrections to
the passive area for these energy-ordered two-particle configurations.

\begin{table}
  \begin{center}
    \begin{tabular}{|c||c|c|c|}
      \hline
      &\multicolumn{3}{c|}{$a_{\JA,R}(\Delta_{12})/\pi R^2$}\\
      & $k_t$ & Cambridge/Aachen & SISCone \\
      \hline
      $0 < \Delta_{12} < R/2$ & $ u(\Delta_{12}/R) $ 
                        & $1$ 
			& $1$	 \\		 
$R/2 < \Delta_{12} < R$ & $u(\Delta_{12}/R) $ 
                        & $w(\Delta_{12}/R) $
			& $1$ \\
$R < \Delta_{12} < 2R$  & $u(\Delta_{12}/R)\; / \; 2 $
                        & $u(\Delta_{12}/R)\; / \; 2 $
			& $u(\Delta_{12}/R)\; / \; 2 $\\
$\Delta_{12} > 2R$      & $1$
                        & $1$
			& $1$ \\
                        \hline
\end{tabular}
\end{center}
\caption{\label{tab:passive-areas} Summary of the passive areas for
  the three jet algorithms for the hard jet in an event containing one
  hard and one soft particle, separated by a $y$-$\phi$  distance
  $\Delta_{12}$. The functions $u$ and $w$ are defined in 
  eqs.~(\ref{eq:u}) and (\ref{eq:w}).}
\end{table}
 
\subsection{Area scaling violation}
\label{sec:area-scal-viol-passive}

Given that jet passive areas are modified by the presence of a soft
particle in the neighbourhood of a jet, one expects the average jet
area to acquire a logarithmic dependence on the jet transverse
momentum when the jets acquire a sub-structure as a consequence of radiative
emission of gluons.
To determine its coefficient we shall work in
the approximation of small jet radii, motivated by the observation
that corrections for finite $R$, proportional to powers of $R$, are
usually accompanied by a small coefficient~\cite{de
  Florian:2007fv,CDMS}.

In the small-angle limit, the QCD matrix element for the emission of
the perturbative soft gluon, $p_2$, is
\begin{equation}
  \label{eq:softcoll}
  \frac{dP}{dp_{t2} \, d\Delta_{12}} = 
  C_1 \frac{2\as(p_{t2} \Delta_{12})}{\pi}
  \frac{1}{\Delta_{12}} \frac{1}{p_{t2}}\,,
\end{equation}
where $C_1$ is $C_F$ or $C_A$ according to whether the hard particle
$p_1$ is a quark or a gluon. We make use of the fact that
$\Delta_{12}$ is just the angle between the two particles (to within a
longitudinal boost-dependent factor), and will assume that $R$ is sufficiently
small that the small-angle approximation is legitimate. The scale of
the coupling is taken to be the transverse momentum of $p_{2}$
relative to $p_1$.

At order $\as$ the mean jet area with a given jet algorithm (\text{JA}) can be
written
\begin{equation}
  \label{eq:ajfr-base}
  \langle a_{\text{JA},R}\rangle 
  = a_{\text{JA},R}(0) + \langle \Delta a_{\text{JA},R}\rangle
  = \pi R^2 + \langle \Delta a_{\text{JA},R}\rangle\,,
\end{equation}
where we explicitly isolate the $\order{\as}$ higher-order contribution,
\begin{equation}
  \label{eq:ajfr}
  \langle \Delta  a_{\text{JA},R}\rangle \simeq \int_0^{2R} d\Delta_{12} 
  \int_{{Q_0}/\Delta_{12}}^{p_{t1}} d p_{t2} 
  \frac{dP}{dp_{t2} \, d\Delta_{12}} (a_{\text{JA},R}(\Delta_{12}) - a_{\text{JA},R}(0))\,,
\end{equation}
with the $- a_{\text{JA},R}(0)$ term accounting for virtual
corrections. $\avg{\cdots}$ represents therefore an average over perturbative
emission. Note that because of the $1/p_{t2}$ soft divergence for
the emission of $p_{t2}$, eq.~(\ref{eq:ajfr}) diverges unless one
specifies a lower limit on $p_{t2}$ --- accordingly we have to
introduce an infrared cutoff ${Q_0}/\Delta_{12}$ on the $p_t$ of the emitted
gluon. This value results from requiring that the transverse momentum
of $p_2$ relative to $p_1$, \ie $p_{t2} \Delta_{12}$, be larger than
${Q_0}$. 
The fact that we need to place a lower limit on $p_{t2}$ means that
jet areas are infrared unsafe\footnote{An exception is for the SISCone
  algorithm with a cut, $p_{t,min}$, on the minimum transverse
  momentum of protojets that enter the split--merge procedure
  procedure --- in that situation ${Q_0}$ is effectively replaced by
  $p_{t,min}$.}  --- they cannot be calculated order by order in
perturbative QCD and for real-life jets they 
will depend on the details of non-perturbative
effects (hadronisation). One can account for this to some extent by
leaving ${Q_0}$ as a free parameter and examining the dependence of
the perturbative result on ${Q_0}$.\footnote{
  One should of course bear in mind that the multi-particle structure
  of the hadron level is such that a single-gluon result cannot
  contain all the relevant physics --- this implies that one should,
  in future work, examine multi-soft gluon radiation as well, perhaps
  along the lines of the calculation of non-global
  logarithms~\cite{Dasgupta:2001sh,Banfi:2002hw,ApplebySeymour,BanfiDasgupta},
  though it is not currently clear how most meaningfully to carry out
  the matching with the non-perturbative regime.
  Despite these various issues, we shall see in
  section~\ref{sec:real-life} that the single-gluon results work
  remarkably well in comparisons to Monte Carlo predictions. In that
  section, we shall also argue that in cases with pileup, the pileup
  introduces a natural semi-hard (\ie perturbative) cutoff scale that
  replaces $Q_0$.}

As concerns the finiteness of the $\Delta_{12}$ integration, all the
jet algorithms we consider have the property that 
\begin{subequations}
  \begin{align}
    \label{eq:lim-delta2zero}
    &\lim_{\Delta_{12}\to 0} a_{\text{JA},R}(\Delta_{12}) = 
     a_{\text{JA},R}(0) = 
    \pi R^2\,;
    \\
    &a_{\text{JA},R}(\Delta_{12}) = \pi R^2\quad \mbox{for $\Delta_{12} >
      2R$}
  \end{align}
\end{subequations}
so that the integral converges for $\Delta_{12} \to 0$, and we can
place the upper limit at $\Delta_{12} = 2R$.

After evaluating eq.~(\ref{eq:ajfr}), with the replacement $\Delta_{12}
\to R$, both in the lower limit of the $p_{t2}$ integral and the
argument of the coupling,
we obtain
\begin{equation}
  \label{eq:delta-ajf-res}
  \langle \Delta a_{\text{JA},R}\rangle = d_{\text{JA},R} \frac{2\as C_1}{\pi} \ln
  \frac{R p_{t1}}{{Q_0}} \;, \qquad
  d_{\text{JA},R} = \int_0^{2R} \frac{d\theta}{\theta} ( a_{\text{JA},R}(\theta) -
  \pi R^2)\,,
\end{equation}
in a fixed coupling approximation, and 
\begin{equation}
  \label{eq:delta-ajf-res-running}
  \langle \Delta a_{\text{JA},R}\rangle = 
  d_{\text{JA},R} 
  \frac{C_1}{\pi b_0}
   \ln \frac{\as({Q_0})}{\as(R p_{t1})}\,,
\end{equation}
with a one-loop running coupling, where $b_0 = \frac{11C_A -
  2n_f}{12\pi}$. The approximation 
$\Delta_{12} \sim R$ in the argument of the running coupling in the
integrand corresponds to neglecting terms of $\order{\as}$
without any enhancements from logarithms of $R$ or $p_{t1}/{Q_0}$.

The results for $d_{\JA,R}$ are
\begin{subequations}
  \begin{align}
    d_{k_t,R} &= \bigg(\frac{\sqrt{3}}{8} + \frac{\pi}{3} + \xi\bigg)
    R^2\,\simeq\,
    0.5638 \, \pi R^2\,,
    \\ 
    d_{\cam,R} &= \bigg(\frac{\sqrt{3}}{8} + \frac{\pi}{3} - 2\xi \bigg) R^2
    \,\simeq\,
    0.07918 \,\pi R^2\,,
    \\
    d_{\cone,R}
    &= \bigg(-\frac{\sqrt{3}}{8}+\frac{\pi }{6} - \xi\bigg) R^2 \,\simeq\, 
    -0.06378 \, \pi R^2\,,
  \end{align}
\end{subequations}
where
\begin{equation}
  \label{eq:xi}
  \xi \equiv
  \frac{\psi'(1/6)+\psi'(1/3)-\psi'(2/3)-\psi'(5/6)}{48\sqrt{3}}
  \,\simeq\, 0.507471\,,
\end{equation}
with $\psi'(x) = d^2/dx^2 (\ln \Gamma(x))$. One notes that the
coefficient for the $k_t$ algorithm is non-negligible, given that it
is multiplied by the quantity $2\as C_1/\pi \ln R p_{t1}/{Q_0}$ in
eq.~(\ref{eq:delta-ajf-res}) (or its running coupling analogue), which
can be of order $1$. In contrast the coefficients for Cambridge/Aachen
and SISCone are much smaller and similar (the latter being however of
opposite sign).  Thus $k_t$ areas will depend significantly more on
the jet $p_t$ than will those for the other algorithms.

The fluctuation of the areas can be calculated in a similar way. Let us
define 
\begin{equation}
  \label{eq:passive-fluct-decomp}
  \langle \sigma^2_{\JA,R} \rangle = \langle a_{\text{JA},R}^2\rangle
  - \langle a_{\text{JA},R}\rangle^2 = \sigma^2_{\JA,R}(0) + \langle
  \Delta \sigma^2_{\JA,R} \rangle\,,\qquad\quad \sigma^2_{\JA,R}(0) = 0\,,
\end{equation}
where we have introduced $\sigma^2_{\JA,R}(0)$, despite its being null,
so as to facilitate comparison with later results. We then evaluate
\begin{equation}
\label{eq:fluct}
\langle
  \Delta \sigma^2_{\JA,R} \rangle=
\langle \Delta a_{\text{JA},R}^2\rangle -
\langle \Delta a_{\text{JA},R}\rangle^2 \simeq \langle \Delta a_{\text{JA},R}^2\rangle \, ,
\end{equation}
where we neglect $\langle \Delta a_{\text{JA},R}\rangle^2$ since it is of 
$\order{\as^2\ln^2(R p_{t1}/{Q_0})}$.
The calculation of $\langle \Delta a_{\text{JA},R}^2\rangle$ proceeds
much as for $\langle \Delta a_{\text{JA},R}\rangle$ and gives
\begin{equation}
  \label{eq:delta-ajf-res2}
  \langle \Delta a_{\text{JA},R}^2\rangle =
  s_{\text{JA},R}^2\frac{C_1}{\pi b_0}
  \ln \frac{\as({Q_0})}{\as(R p_{t1})}\;, \qquad
  s_{\text{JA},R}^2 = \int_0^{2R} \frac{d\theta}{\theta} ( a_{\text{JA},R}(\theta) -
  \pi R^2)^2\,
\end{equation}
for running coupling.
The results are
\begin{subequations}\label{eq:fluct_passive_coefs}
  \begin{align}
    s_{k_t,R}^2 &=  \bigg( \frac{\sqrt{3}\pi}{4} - \frac{19}{64}
    -\frac{15\zeta(3)}{8} + 2\pi \xi \bigg) R^4 \,\simeq\, 
    (0.4499 \, \pi R^2)^2\,,
    \\
    s_{\cam,R}^2 &=  \bigg( \frac{\sqrt{3}\pi}{6} -\frac{3}{64} -
    \frac{\pi^2}{9} -\frac{13\zeta(3)}{12} +\frac{4\pi}{3} \xi
    \bigg) R^4 \,\simeq\, 
    (0.2438\, \pi R^2)^2\,,
    \\
    s_{\cone,R}^2
    &=  \bigg( \frac{\sqrt{3}\pi}{12} -\frac{15}{64} -\frac{\pi^2 }{18}
    -\frac{13\zeta(3)}{24} +\frac{2\pi}{3} \xi \bigg) R^4 \,\simeq\, 
    (0.09142\, \pi R^2)^2\,.
  \end{align}
\end{subequations}
We have a hierarchy between algorithms that is similar to that observed for
the average area scaling violations, though now the coefficient for
Cambridge/Aachen is more intermediate between the other two.

\subsection[$n$-particle properties and the Voronoi area]{$\boldsymbol n$-particle properties and the Voronoi area}
\label{sec:n-particle-passive}

The only algorithm for which one can make any statement about the
passive area for a general $n$-particle configuration is the $k_t$
algorithm.

Because of the $k_t$ distance measure, the single ghost will cluster
with one of the event particles before any other clustering takes
place. 
One can determine the region in which the ghost will cluster with a
given particle, and this is a definition of the area $a_{\kt,R}(p_i)$
of a particle $p_i$.  Since the ghost-particle clustering will occur
before any particle-particle clustering, the jet area will be the sum
of the areas of all its constituent particles:
\begin{equation}
  \label{eq:particle-areas}
  a_{\kt,R}(J) = \sum_{p_i \in J} a_{\kt,R}(p_i)\,.
\end{equation}

Can anything be said about the area of a particle? The ghost will
cluster with the event particle to which it is closest, as long it is
within a distance $R$. There exists a geometrical construction known
as the Voronoi diagram, which subdivides the plane with a set of
vertices into cells around each vertices.\footnote{It is this same
  geometrical construction that was used to obtain a nearest neighbour
  graph that allowed $k_t$ jet clustering to be carried out in $N \ln
  N$ time~\cite{fastjet}.} %
Each cell has the property that all points in the cell have as their
closest vertex the cell's vertex.  Thus the Voronoi cell is remarkably
similar to the region in which a ghost will cluster with a particle.
The only difference arises because of the limitation that the ghost
should be within a distance $R$ of the particle --- this causes the
area of particle $i$ to be the area of its Voronoi cell ${\cal V}_i$
intersected with a circle of radius $R$, ${\cal C}_{i,R}$, centred on
the particle. This leads us to define a Voronoi area for a particle,
$a_R^{\cV}(p_i)$,
\begin{equation}
  \label{eq:voronoi-intersect}
  a_R^{\cV}(p_i) \equiv \text{area}({\cal V}_i \cap {\cal
    C}_{i,R})\,.
\end{equation}
Thus given a set of momenta, the passive area of a $k_t$ jet can be
directly determined from the Voronoi diagram of the
event,\footnote{Strictly speaking it should be the Voronoi diagram on
  a $y-\phi$ cylinder, however this is just a technical detail.} using
eq.~(\ref{eq:particle-areas}) and the relation
\begin{equation}
  \label{eq:kt-voronoi}
  a_{\kt,R}(p_i) = a_R^{\cV}(p_i)\,.
\end{equation}
\begin{figure}
  \centering
  \includegraphics[width=0.7\textwidth]{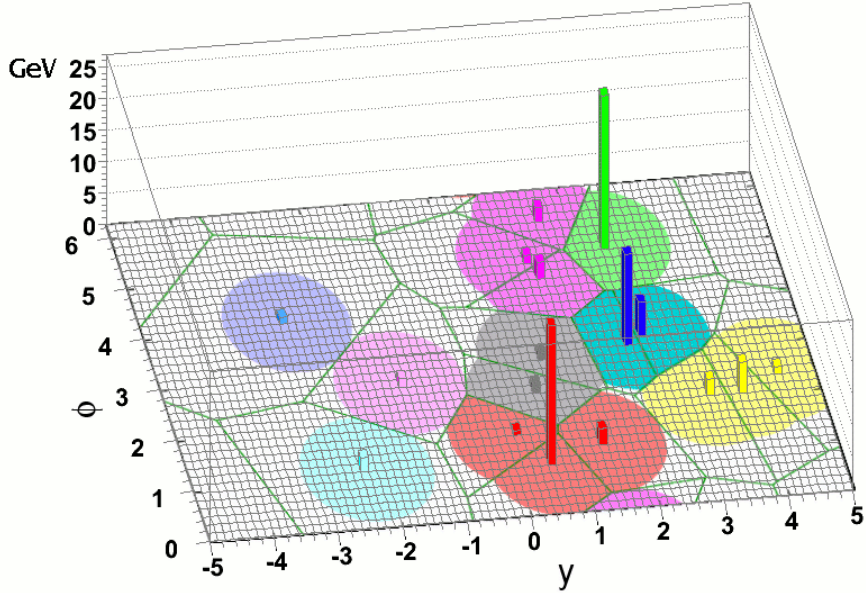}
  \caption{The passive area of jets in a parton-level event generated
    by Herwig and clustered with the $k_t$ algorithm with $R=1$. The
    towers represent calorimeter cells containing the particles, the
    straight (green) lines are the edges of the Voronoi cells and the
    shaded regions are the areas of the jets.}
  \label{fig:event-voronoi-mod}
\end{figure}%
It is quite straightforward to see that this result holds for the
2-particle case, because the Voronoi diagram there consists of a
single line, equidistant between the two points. It divides the plane
into two half-planes, each of which is the Voronoi cell of one of the
particles (this is best seen in fig.~\ref{fig:overlaps}c). The
intersection of the halfplane with the circle of radius $R$ centred on
the particle has area ${\scriptstyle{\frac{1}{2}}}\pi R^2
u(\Delta_{12}/R)$, and this immediately gives us the results
eqs.~(\ref{eq:a_kt_Dlt1}), (\ref{eq:a_kt_Dgt1}) according to whether
the particles cluster into a single jet or not.

The Voronoi construction of the $k_t$-algorithm passive area is
illustrated for a more complex event in
fig.~\ref{fig:event-voronoi-mod}. One sees both the Voronoi cells and
how their intersection with circles of radius $R=1$ gives the area of
the particles making up those jets.

Note that it is not possible to write passive areas for jet algorithms
other than $k_t$ in the form eq.~(\ref{eq:particle-areas}). One can
however introduce a new type of area for a generic algorithm, a
{\it Voronoi area}, in the form 
\begin{equation}
  \label{eq:jet-voronoi-area}
  a_{\JA,R}^{\cV}(J) = \sum_{p_i \in J} a_{R}^{\cV}(p_i)\,.
\end{equation}
While for algorithms other than $k_t$ (for which, as we have seen,
$a_{\kt,R}(J) = a_{\kt,R}^{\cV}(J)$), this area is not in general
related to the clustering of any specific kind of background
radiation, it can nevertheless be a useful tool, because its numerical
evaluation is efficient~\cite{Fortune,fastjet} and as we shall discuss
later (sec.~\ref{sec:n-particle-active}), for dense events its value
coincides with both passive and active area definitions.

\section{Active Area}
\label{sec:active}

To define an active area,
as for the passive area, we start with an
event composed of a set of particles $\{p_i\}$ which are clustered
into a set of jets $\{J_i\}$ by some infrared-safe jet algorithm.
However, instead of adding a single soft ghost particle, we now add a
dense coverage of ghost particles, $\{g_i\}$, randomly distributed in
rapidity and azimuth, and each with an infinitesimal transverse
momentum.\footnote{In most cases the distribution of those transverse
  momenta will be irrelevant, at least in the limit in which the
  density of ghosts is sufficiently high.} %
The clustering is then repeated including the set of particles plus
ghosts. 

During the clustering the ghosts may cluster both with each
other and with the hard particles. This more `active' participation in
the clustering is the origin of the name that we give to the area to
be defined shortly. It contrasts with the definition of
section~\ref{sec:passive}, in which the single ghost acted more as a
passive spectator, and in particular could not cluster with any other
ghosts (there weren't any).

\begin{figure}
  \centering
  \includegraphics[width=0.7\textwidth]{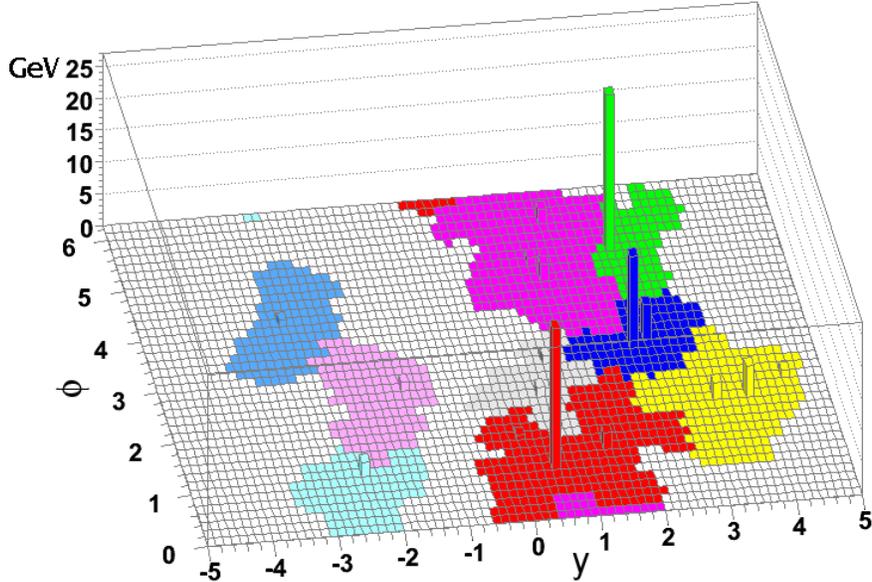}
  \caption{Active area for the same event as in
    figure~\ref{fig:event-voronoi-mod}, once again clustered with the
    $k_t$ algorithm and $R=1$. Only the areas of the hard jets have
    been shaded --- the pure `ghost' jets are not shown.}
  \label{fig:example-active-area}
\end{figure}

Because of the infrared safety of any proper jet algorithm, even the addition of
many ghosts does not change the momenta of the final jets $\{J_i\}$.
However these jets do contain extra particles, ghosts, and we use the number
of ghosts in a jet as a measure of its area. Specifically, if the
number of ghosts per unit area (on the rapidity-azimuth cylinder) is
$\nu_g$ and $\cNg(J)$ is the number of ghosts contained in jet $J$,
then the (scalar) active area of a jet, given the specific set of ghosts
$\{g_i\}$ is 
\begin{equation}
  \label{eq:act_area_given_gi}
  A (J \,|\, {\{g_i\}})  =   \frac{\cNg(J)}{\nu_g} \,.
\end{equation}
An example of jet areas obtained in this way is shown in
figure~\ref{fig:example-active-area}. One notes that the boundaries of
the jets are rather ragged. Clustering with a different set of ghosts
would lead to different boundaries.
This is because the ghosts can cluster among themselves to form
macroscopic subjets, whose outlines inevitably depend on the specific
set of initial ghosts,\footnote{This is a form of dynamic
  magnification of the microscopic local breaking of translational
  invariance introduced by the ghosts' randomness.} and these then
subsequently cluster with true event particles. This can happen for
any density of ghosts, and thus the jet boundaries tend, for most
algorithms, to be sensitive to the randomness of the initial sets of
ghosts.

%
This randomness propagates through to the number of ghosts clustered
within a given jet, even in the limit $\nu_g \to \infty$, resulting in
a different area each time. To obtain a unique answer for active area
of a given jet one must therefore average over many sets of ghosts, in
addition to taking the limit of infinite ghost density,\footnote{One
  may wonder if the averaged area (and its dispersion) depends on the
  specific nature of the fluctuations in ghost positions and momenta
  across ensembles of ghosts --- for a range of choices of these
  fluctuations, no significant difference has been observed (except in
  the case of pure ghost jets with SISCone, whose split--merge step
  introduces a strong dependence on the microscopic event structure).
}
\begin{equation}
  \label{eq:act_area}
  A(J) = \lim_{\nu_g \to \infty} \left \langle
  A (J \,|\, {\{g_i\}}) \right \rangle_g \,.
\end{equation}
Note that as one takes $\nu_g \to \infty$, the ghost transverse
momentum density, $\nu_g \langle g_t \rangle$, is to be kept infinitesimal.

The active area should  bear a close resemblance to the average
susceptibility of the jet to a high density of soft radiation (\eg
minimum-bias pileup), since the many soft particles will cluster
between each other and into jets much in the same way as will the
ghosts.

One may also define the standard deviation $\Sigma(J)$ of the
distribution of a jet's active area across many ghost ensembles,
\begin{equation}
  \label{eq:act_area_stddev}
  \Sigma^2(J) = \lim_{\nu_g \to \infty} \left \langle
  A^2 (J \,|\, {\{g_i\}}) \right \rangle_g  - A^2(J)\,.
\end{equation}
This provides a measure of the variability of a given jet's
contamination from (say) pileup and is closely connected with the
momentum resolution that can be obtained with a given jet algorithm.

A feature that arises when adding many ghosts to an event is that some
of the final jets contain nothing but ghost particles.  They did not
appear in the original list of $\{J_i\}$ and we refer to them as pure
ghost jets. These `ghost' jets (not shown in
fig.~\ref{fig:example-active-area}), fill all of the `empty' area, at
least in jet algorithms for which all particles are clustered into
jets.
They will be similar to the jets formed from purely soft radiation in
events with minimum-bias pileup, and so are interesting to study
in their own right.

We can also define a 4-vector version of the active area (in analogy
with the 4-vector passive area). It is given by 
\begin{equation}
  \label{eq:act4_area_given_gi}
  A_\mu (J \,|\, {\{g_i\}})  =   
  \frac{1}{\nu_g \langle g_t \rangle } \sum_{g_i \in J} g_{\mu i} \,,
  \qquad\quad
  A_\mu(J) = \lim_{\nu_g \to \infty} \left \langle
  A_\mu (J \,|\, {\{g_i\}}) \right \rangle_g \,.
\end{equation}
The sum of the $g_{\mu i}$ is to be understood as carried out in the same
recombination scheme as used in the jet clustering.%
\footnote{Though we do not give the details it is simple to extend the
  4-vector active area definition to hold also for a general IR safe
  jet algorithm, in analogy with the extension of the passive area
  definition in eq.~(\ref{eq:passive-4vect-area-gen}).}

\subsection{Areas for 1-particle  configurations and for ghost jets}
\label{sec:active_1point}

\subsubsection{$\boldsymbol{\kt}$ and Cambridge/Aachen}\label{sec:active_1point_kt_cam}

The active area for the $k_t$ and Cambridge/Aachen algorithms is most
readily studied numerically, by directly clustering large numbers of
ghost particles, possibly together with one or more hard particles.
This is feasible because of the availability of fast computational
methods for carrying out the clustering in these algorithms,
implemented in the FastJet package~\cite{fastjet}. Typically we add
ghosts with a density $\nu_g$ of $\sim 100$ per unit area,\footnote{They are
  placed on a randomly scattered grid, in order to limit the impact of
  the finite density, \ie one effectively carries out quasi Monte
  Carlo integration of the ghost ensembles, so that that finite
  density effects ought to vanish as $\nu_g^{-3/4}$, rather than $\nu_g^{-1/2}$
  as would be obtained with completely random placement. }
in the rapidity region
$|y| < y_{max} = 6$, and study jets in the region $|y| <
y_{max}-R$. This leads to about 7500 ghost particles, which can be
clustered in about $0.1\,\mathrm{s}$ on a $3.4\,\mathrm{GHz}$
processor. Each ghost is given a transverse momentum $\sim
10^{-100}\GeV$ and the one hard particle that we study has a
transverse momentum of $100\GeV$.
The results are insensitive to the values chosen as long as their
ratio is sufficiently large.
We investigate in Appendix \ref{sec:transition} how the distribution
of the ``1-hard-parton'' jet area gets modified when the transverse
momentum of the parton is progressively reduced below the scale of a
generic set of soft particles.

\begin{figure}
  \centering
  \includegraphics[width=0.48\textwidth]{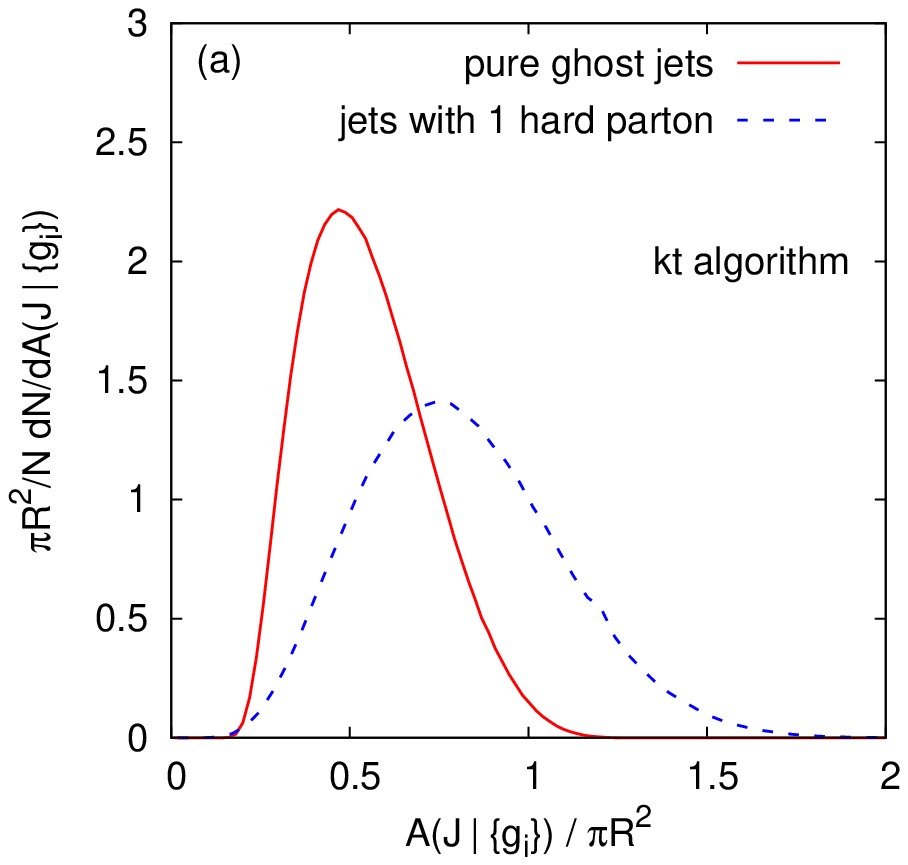}\hfill
  \includegraphics[width=0.48\textwidth]{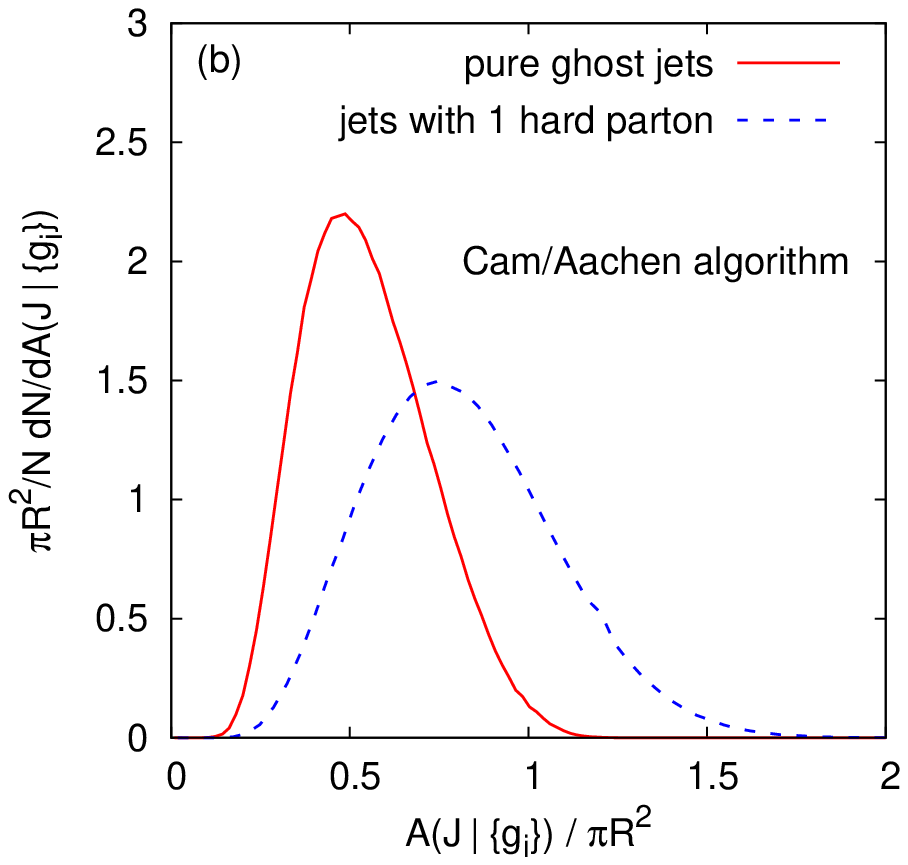}
  \caption{Distribution of active areas for pure ghost jets and
    jets with a single hard particle: (a) $k_t$ algorithm, (b)
    Cambridge/Aachen algorithm. }
  \label{fig:ghost-areas}
\end{figure}

Figure~\ref{fig:ghost-areas} shows the distribution of values of $A (J
\,|\, {\{g_i\}})$ for pure ghost jets and jets with one hard particle.
The distribution is obtained over a large ensemble of sets of
ghosts.\footnote{In this particular case we have used about $10^7$
  separate random ghost sets, in order to obtain a smooth curve for
  the whole distribution. When calculating areas in physical events
  (or even at parton-shower level) the multiple real particles in the
  jet ``fix'' most of the area, and between 1 and 5 sets of ghosts
  particles are usually sufficient to obtain reliable area results
  (this is the case also for SISCone).}
Let us concentrate initially on the case with a hard particle. Firstly
the average active area, eq.~(\ref{eq:act_area}) differs noticeably
from the passive result of $\pi R^2$:
\begin{subequations}
\label{eq:area-hard-particle}
  \begin{align}
    A_{k_t,R}  (\text{one-particle-jet}) &\,\simeq\, 0.812 \,\pi R^2\,, \\
    A_{\cam,R} (\text{one-particle-jet}) &\,\simeq\, 0.814 \,\pi R^2\,.
  \end{align}
\end{subequations}
Secondly, the distributions of the area in fig.~\ref{fig:ghost-areas}
are rather broad. The randomness in the initial distribution of ghosts
propagates all the way into the shape of the final jet and hence its
area. This occurs because the $k_t$ and Cambridge/Aachen algorithms
flexibly adapt themselves to local structure (a good property when
trying to reconstruct perturbative showering), and once a random
perturbation has formed in the density of ghosts this seeds further
growth of the soft part of the jet.
The standard deviations of the resulting distributions are
\begin{subequations}
  \label{eq:ktcamsigmas}
  \begin{align}
    \Sigma_{k_t,R}  (\text{one-particle-jet}) &\,\simeq\, 0.277 \,\pi R^2\,, \\
    \Sigma_{\cam,R} (\text{one-particle-jet}) &\,\simeq\, 0.261 \,\pi R^2\,.
  \end{align}
\end{subequations}

Figure~\ref{fig:ghost-areas} also shows the distribution of areas for
pure ghost jets. One sees that pure ghost jets typically have a
smaller area than hard-particle jets:\footnote{Obtaining these values
  actually requires going beyond the ghost density and the rapidity
  range previously mentioned. In fact, when going to higher accuracy
  one notices the presence of small edge and finite-density effects,
  ${\cal O}(R/(y_{max}-R))$ and ${\cal O}(1/(\nu_g R^2))$ to 
  some given power.  Choosing
  the ghost area sufficiently small to ensure that finite-density
  effects are limited to the fourth decimal (in practice this
  means $1/(\nu_g R^2) < 0.01$) and extrapolating to infinite
  $y_{max}$ one finds
\begin{subequations}
  \begin{align}
    A_{k_t}  (\text{ghost-jet}) &\,\simeq\, (0.5535 \pm 0.0005) \,\pi R^2\,, \\
    A_{\cam} (\text{ghost-jet}) &\,\simeq\, (0.5505 \pm 0.0005) \,\pi R^2\,,
  \end{align}
\end{subequations}  
with a conservative estimate of the residual uncertainty.
This points to a small but statistically significant 
difference between the two algorithms.
}
\begin{subequations}
\label{eq:area-pure-ghost}
  \begin{align}
    A_{k_t,R}  (\text{ghost-jet}) &\,\simeq\, 0.554 \,\pi R^2\,, \\
    A_{\cam,R} (\text{ghost-jet}) &\,\simeq\, 0.551 \,\pi R^2\,,
  \end{align}
\end{subequations}
and the standard deviations are
\begin{subequations}
\label{eq:area-pure-ghost-stddev}
  \begin{align}
    \Sigma_{k_t,R}  (\text{ghost-jet}) &\,\simeq\, 0.174 \,\pi R^2\,, \\
    \Sigma_{\cam,R} (\text{ghost-jet}) &\,\simeq\, 0.176 \,\pi R^2\,.
  \end{align}
\end{subequations}
The fact that pure ghost jets are smaller than hard jets has an
implication for certain physics studies: one expects jets made of soft
`junk' (minimum bias, pileup, thermal noise in heavy ions) to have
area properties similar to ghost jets; since they are smaller on
average than true hard jets, the hard jets will emerge from the junk
somewhat more clearly than if both had the same area.

\subsubsection{SISCone}\label{sec:active_1point_cone}

The SISCone algorithm is unique among the algorithms studied here in
that its active area is amenable to analytical treatment, at least in
some cases.

\begin{figure}
  \centering
  \includegraphics[width=\textwidth]{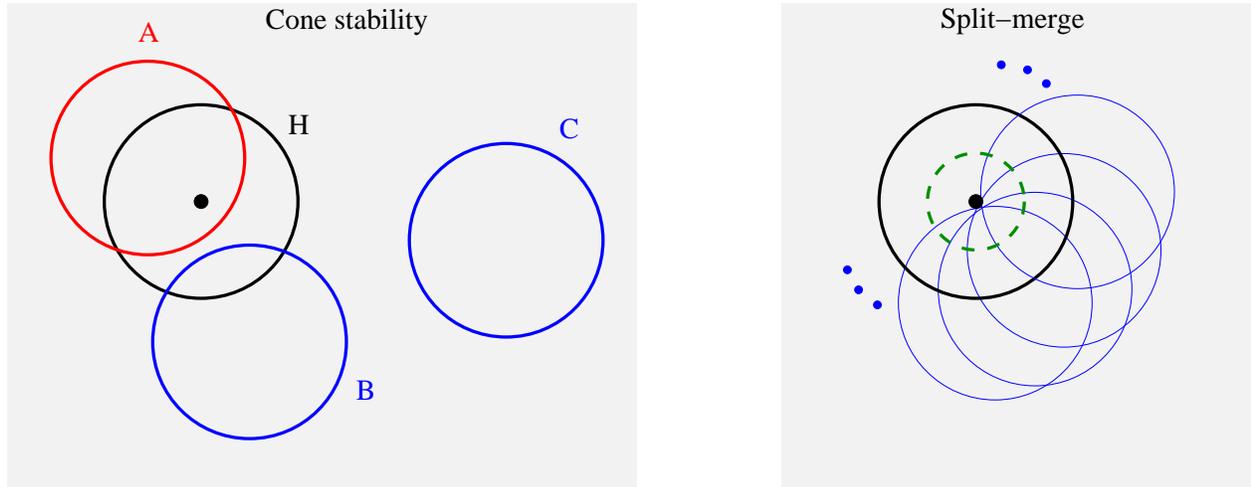}
  \caption{Left: the hard particle with the stable cone (H) centred
    on it, an example of a cone (A) that is unstable because it also
    contains the hard particle, and of two cones (B) and (C) that
    contain just ghost particles and are therefore stable. Right: some
    of the stable ghost cones (thin blue circles) that have the
    largest possible overlap with (H), together with the boundary of
    the hard jet after the split--merge procedure (dashed green line).
    In both diagrams, the grey background represents the uniform
    coverage of ghosts.  }
  \label{fig:cone-active-diag}
\end{figure}

We recall that a modern cone algorithm starts by finding all stable cones.
One stable cone is centred on the single hard particle. Additionally,
there will be a large number of other stable cones, of order of the
number of ghost particles added to the event~\cite{siscone}. In the
limit of an infinite number of ghosts, all cones that can be drawn in
the rapidity-azimuth plane and that do not overlap with the hard particle
will be stable. Many of these stable cones will still overlap with the
cone centred on the hard particle, as long as they do not contain
the hard particle itself (see figure~\ref{fig:cone-active-diag}, left).

Next, the SISCone algorithm involves a split--merge procedure. One defines 
$\tilde p_t$ for a jet as  the scalar sum of the transverse momenta of its 
constituents. 
During the split--merge step, SISCone
finds the stable cone with the highest $\tilde p_t$, and then the
next hardest stable cone that overlaps with it. To decide whether
these two cones (protojets) are to be merged or split, it determines
the fraction of the softer cone's $\tilde p_t$ that is shared with the harder
cone. If this fraction is smaller than some value $f$ (the overlap
parameter of the cone algorithm), the two protojets are split:
particles that are shared between them are assigned to the protojet to
whose centre\footnote{This centre is given by the sum of
  momenta in the protojet before the split--merge operation.} they
are closer.  Otherwise they are merged into a single protojet. This
procedure is repeated until the hard protojet no longer has any
overlap with other protojets. At this point it is called a jet, and
the split--merge procedure continues on the remaining protojets
(without affecting the area of the hard jet).

The maximum possible overlap fraction,\footnote{The fraction of
  momentum coincides with the fraction of area because the ghosts have
  uniform transverse momentum density.} $f_{\text{max}}$, between the
hard protojet and a ghost protojet occurs in the situation depicted in
figure~\ref{fig:cone-active-diag} (right), \ie when the ghost protojet's
centre is just outside the edge of the original hard stable cone (H).
It is given by $f_{\text{max}} =
2-u(1)=\frac{2}{3}-\frac{\sqrt{3}}{2\pi}\approx 0.391$. This means
that for a split--merge parameter $f>f_{\text{max}}$ (commonly used
values are $f = 0.5$ and $f = 0.75$) every overlap between the hard
protojet and a pure-ghost stable cone will lead to a splitting. Since
these pure-ghost stable cones are centred at distances $d > R$
from the hard particle, these splittings will reduce the hard jet to a
circle of radius $R/2$ (the dashed green line in the right hand part
of figure~\ref{fig:cone-active-diag}). The active area of the hard jet
is thus
\begin{equation}\label{eq:active_1point_cone}
A_{\cone,R}(\text{one-particle-jet}) = \frac{\pi R^2}{4} \,.
\end{equation}
This result has been verified numerically using the same technique
employed above for $k_t$ and Cambridge/Aachen.

Note that this area differs considerably from the passive area,
$\pi R^2$. This shows that the cone area is very sensitive to the
structure of the event, and it certainly does not always coincide with
the naive geometrical expectation $\pi R^2$, contrary to assumptions
sometimes made in the literature (see for
example~\cite{Bhatti:2005ai}).

We further note that in contrast to $k_t$ and Cambridge/Aachen
algorithms, the SISCone algorithm always has the same active area for a
single hard particle, independently of fluctuations of an infinitely
dense set of ghosts, \ie
\begin{equation}\label{eq:active_1point_cone_sigma}
  \Sigma_{\cone,R}(\text{one-particle-jet}) = 0 \,.
\end{equation}

\begin{figure}
  \centering
  \includegraphics[height=0.47\textwidth]{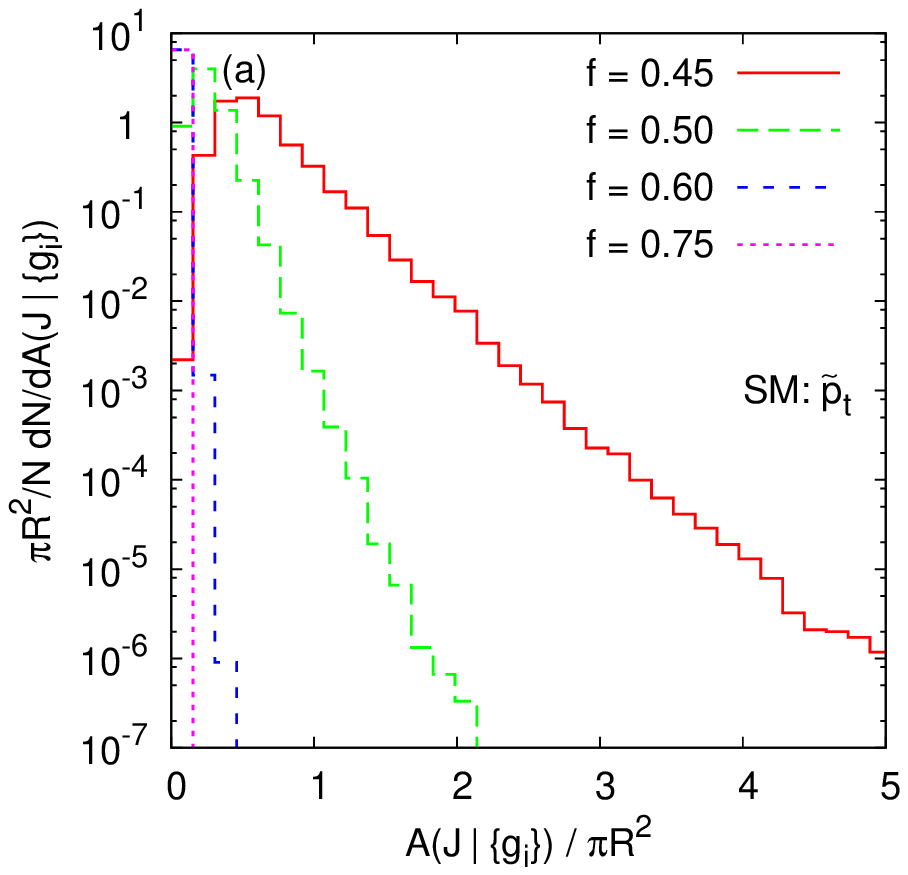}\hfill
  \includegraphics[height=0.47\textwidth]{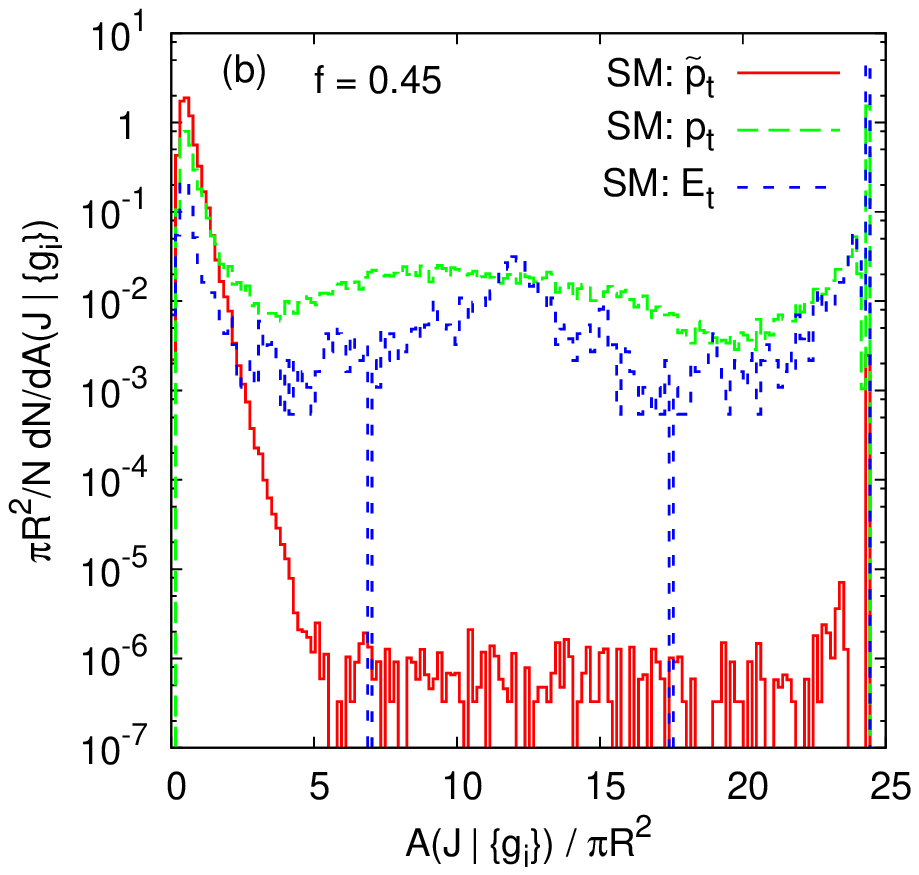}
  \caption{Distribution of pure-ghost jet areas for the SISCone algorithm,
    (a) with different values of the split--merge parameter $f$ and
    (b) different choices for the scale used in the split--merge
    procedure.
    %
    Ghosts are placed on a grid up to $|y|<6$, with an average spacing
    of $0.2\times 0.2$ in $y$, $\phi$, and a random displacement of up
    to $\pm 0.1$ in each direction, and transverse momentum values
    that are uniform modulo a $\pm 5\%$ random rescaling for each
    ghost. We consider all jets with $y < 5$. All jet definitions use
    $R=1$ and multiple passes.}
  \label{fig:siscone-ghost-areas}
\end{figure}

\paragraph{SISCone ghost-jet areas.}
While the area of a hard particle jet could be treated analytically,
this is not the case for the pure-ghost jet area. Furthermore,
numerical investigations reveal that the pure-ghost
area distribution has a much more complicated behaviour than for
$k_t$ or Cambridge/Aachen. One aspect is that the distribution of
pure-ghost jet areas is sensitive to the fine details of how
the ghosts are distributed (density and transverse momentum
fluctuations). Another is that it depends significantly on the details
of the split--merge procedure. Figure~\ref{fig:siscone-ghost-areas}a
shows the distribution of areas of ghost jets for SISCone, for
different values of the split--merge overlap threshold $f$. One sees,
for example,
that for smaller values of $f$ there are occasional rather large ghost
jets, whereas for $f\gtrsim 0.6$ nearly all ghost jets have very small
areas.

One of the characteristics of SISCone that differs from previous cone
algorithms is the specific ordering and comparison variable used to
determine splitting and merging. As explained above, 
the choice made in SISCone was $\tilde
p_t$, the scalar sum of transverse momenta of all particles in a jet.
Previous cone algorithms used either the vector sum of constituent
transverse momenta, $p_t$ (an infrared unsafe choice), or the
transverse energy $E_t = E \sin \theta$ (in a 4-vector recombination
scheme). With both of these choices of variable, split--merge
thresholds $f\lesssim 0.55$ can lead to the formation of `monster'
ghost jets, which can even wrap around the whole cylindrical phase
space.  For $f=0.45$ this is a quite frequent occurrence, as
illustrated in figure~\ref{fig:siscone-ghost-areas}b, where one sees a
substantial number of jets occupying the whole of the phase space (\ie
an area $4\pi y_{\max} \simeq 24 \pi R^2$). Monster jets can be formed
also with the $\tilde p_t$ choice, though it is a somewhat rarer
occurrence --- happening in `only' $\sim 5\%$ of events.\footnote{This
  figure is not immediately deducible from
  fig.~\ref{fig:siscone-ghost-areas}b, which shows results normalised
  to the total number of ghost jets, rather than to the number of
  events.}

We have observed the formation of such monster jets also from normal
pileup momenta simulated with Pythia~\cite{Pythia}, indicating that
this disturbing characteristic is not merely an artefact related to
our particular choice of ghosts. This indicates that a proper choice
of the split--merge variable and threshold is critical in
high-luminosity environments. The results from
Figure~\ref{fig:siscone-ghost-areas}a, suggest that if one wants to
avoid monster jets, one has to choose a large enough value for
$f$. Our recommendation is to adopt $f=0.75$ as a default value for
the split--merge threshold (together with the use of the $\tilde p_t$
variable, already the default in SISCone, for reasons related to
infrared safety and longitudinal boost invariance).

\subsection{Areas for 2-particle configurations}
\label{sec:active_2point}

In this section we study the same problem described in
section~\ref{sec:passive-area-2particle}, \ie the area of jets
containing two particles, a hard one and a softer (but still
``perturbative'') one, \ie eq.~(\ref{eq:strord}), but now for active
areas. As before, the results will then serve as an input in
understanding the dependence of the active area on the jet's
transverse momentum when accounting for perturbative radiation.

\subsubsection{$\boldsymbol{k_t}$ and
  Cambridge/Aachen}\label{sec:active_2point_kt_cam}

As was the case for the active area of a jet containing a single hard
particle, we again have to resort to numerical analyses to study that
of jets with two energy-ordered particles. We define
$A_{\JA,R}(\Delta_{12})$ to be the active area for the energy-ordered
two particle configuration already discussed in section
\ref{sec:passive-area-2particle}. 

Additionally since we have a distribution of areas for the
single-particle active area case, it becomes of interest to study also
$\Sigma_{\JA,R}(\Delta_{12})$ the standard deviation of the
distribution of areas obtained for the two-particle configuration.

\begin{figure}[t]
  \centering
  \includegraphics[width=0.7\textwidth]{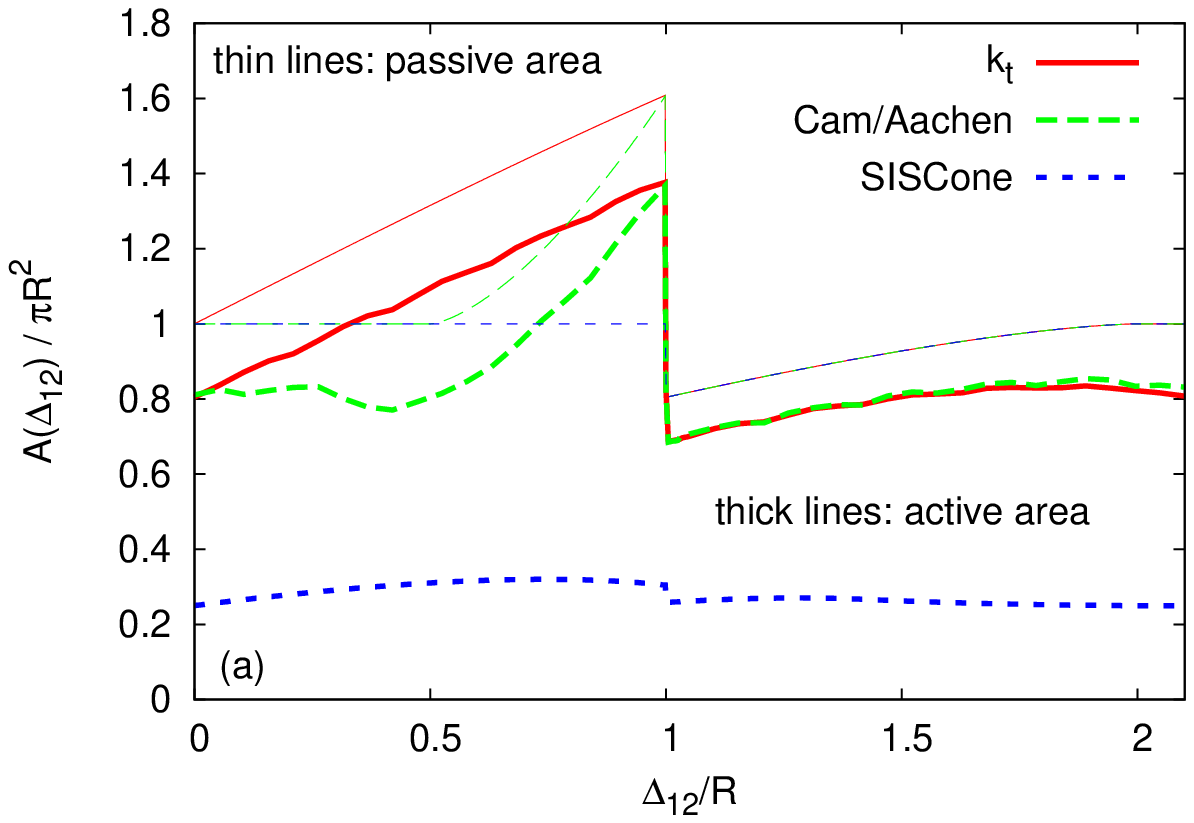}
  \includegraphics[width=0.7\textwidth]{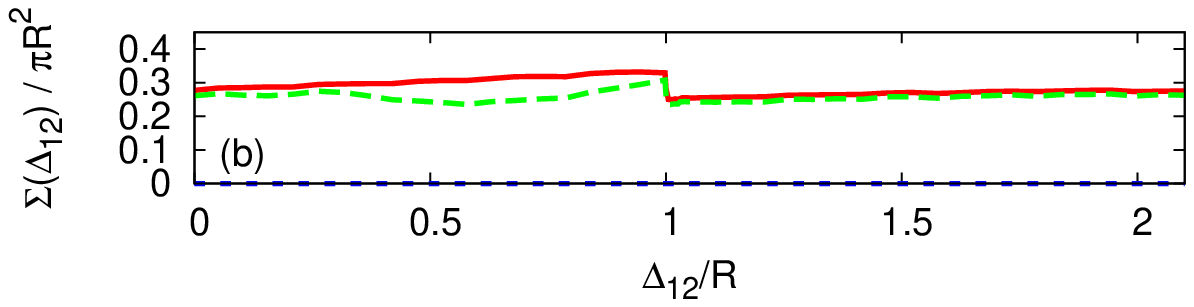}
  \caption{(a) Active areas (divided by $\pi R^2$) for the three jet
    algorithms as a function of the separation between a hard and a
    softer particle. For comparison we also include the passive areas,
    previously shown in fig.~\ref{fig:2point}. (b) The corresponding
    standard deviations.}
  \label{fig:2point-areas-active-3alg}
\end{figure}

The results are shown in figure~\ref{fig:2point-areas-active-3alg}:
the active areas can be seen to be consistently smaller than the
passive ones, as was the case for the 1-particle area, but retain the
same dependence on the angular separation between the two particles.
Among the various features, one can also observe
that the active area does not quite reach the single-particle value
($\simeq 0.81 \pi R^2$) at $\Delta_{12}=2R$, but only somewhat beyond
$2R$.
This contrasts with the behaviour of the passive area.
The figure also shows results for the cone area, discussed in the
following subsection. 

\subsubsection{SISCone}\label{sec:active_2point_cone}

In the case of the SISCone algorithm it is possible to find an analytical
result for the two-particle active area, in an extension of what was
done for one-particle case.\footnote{This is only possible for
  configurations with strong energy ordering between all particles ---
  as soon as 2 or more particles have commensurate transverse momenta
  then the cone's split--merge procedure will include `merge' steps,
  whose effects on the active area are currently beyond analytical
  understanding.}

The stable-cone search will find one or two ``hard'' stable cones: the
first centred on the hard particle and the second centred on the soft
one, present only for $\Delta_{12}>R$. The pure-ghost stable cones
will be centred at all positions distant by more than $R$ from both
$p_1$ and $p_2$, \ie outside the two circles centred on $p_1$ and
$p_2$.

We shall consider the active area of the jet centred on the hard particle
$p_1$. When $\Delta_{12}>R$, the jet centred on the soft particle has
the same area.

\begin{figure}
\includegraphics[scale=0.50]{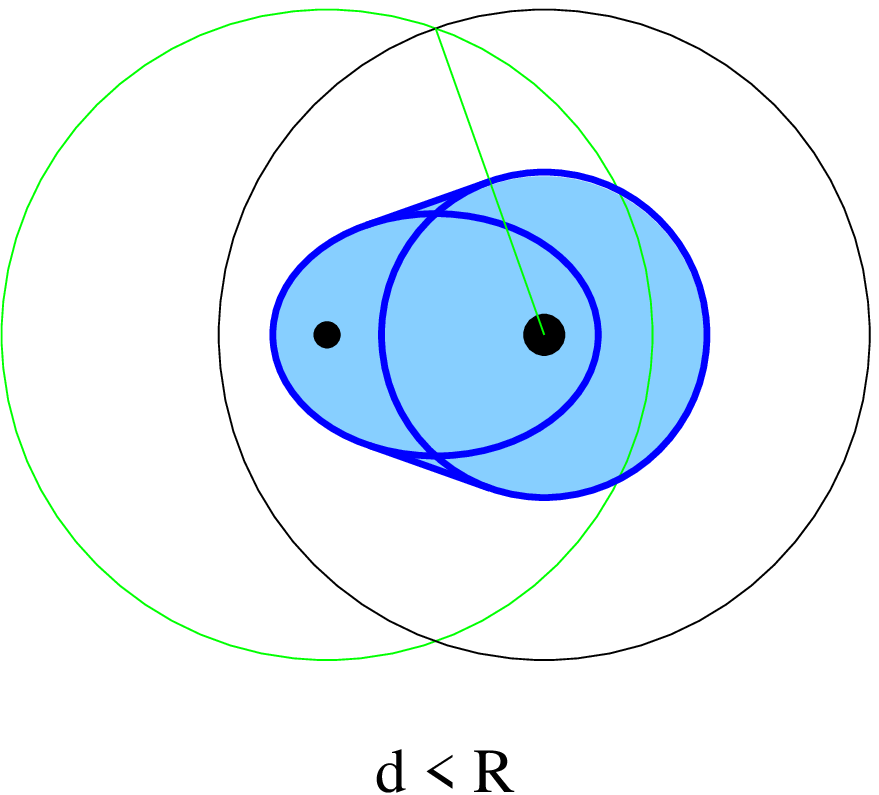}\;\;
\includegraphics[scale=0.50]{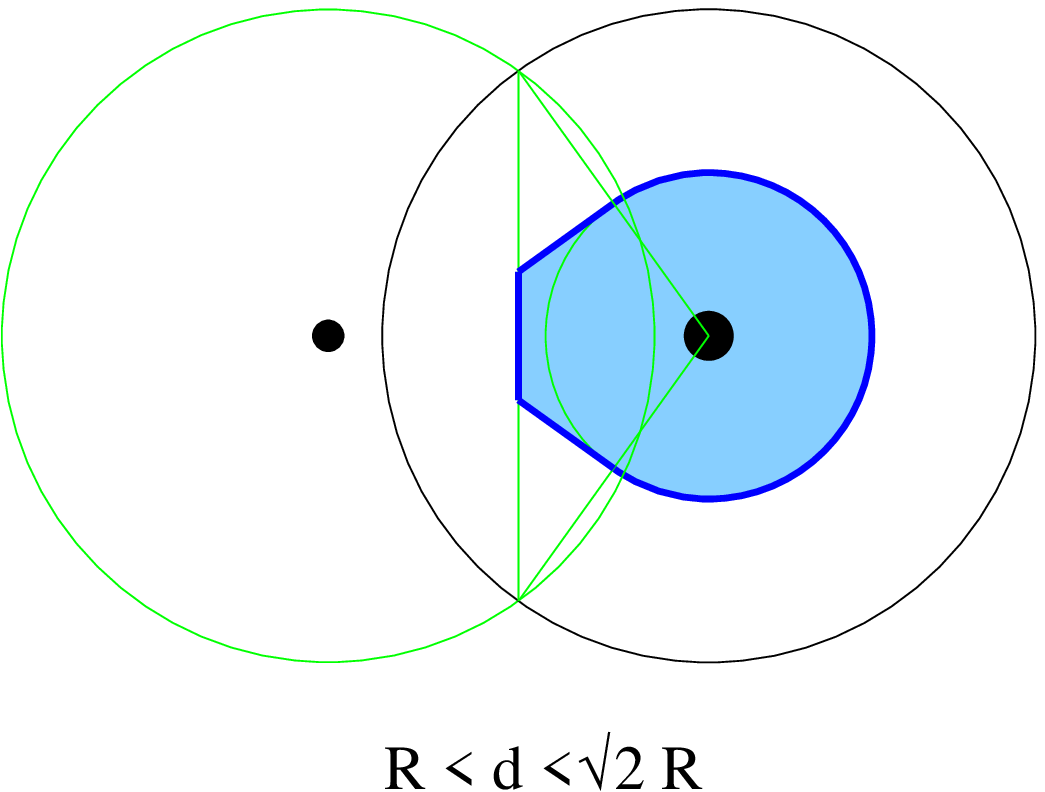}\;\;
\includegraphics[scale=0.50]{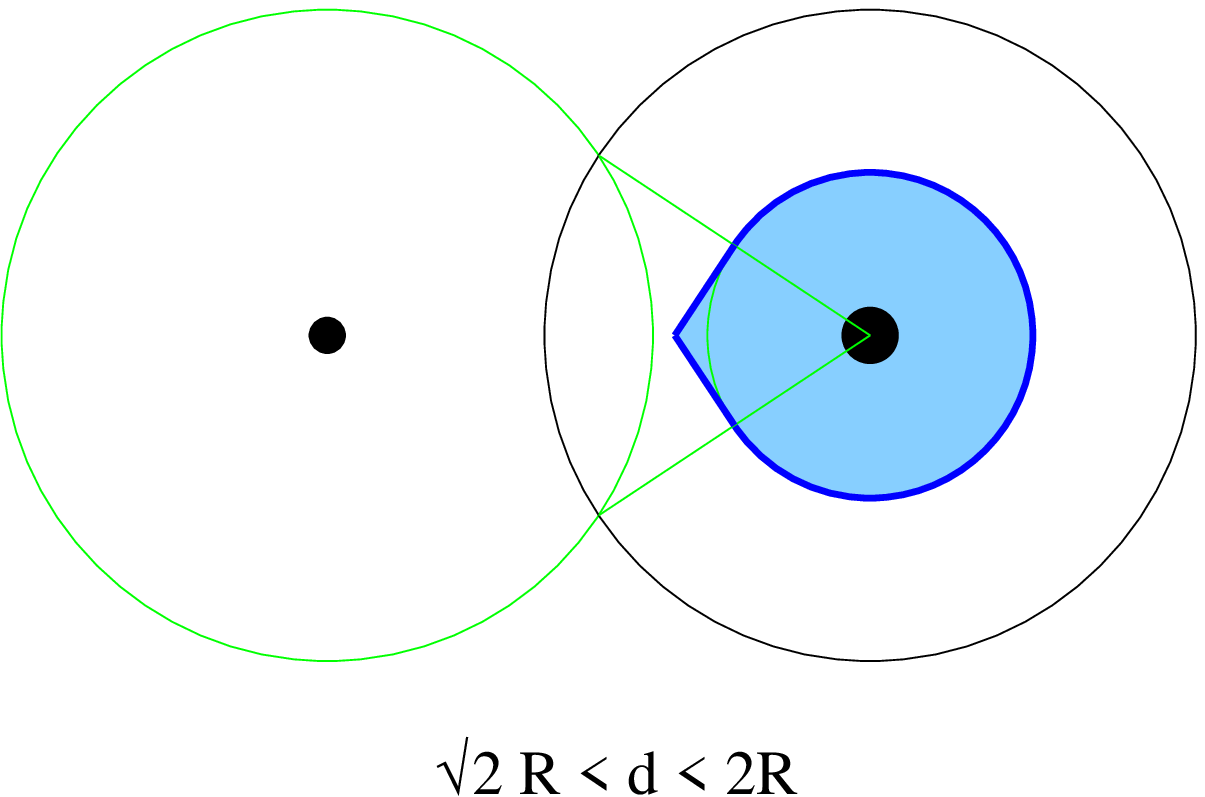}
\caption{Picture of the active jet area for 2-particle configurations
  in the case of the SISCone jet algorithm. The black points represent
  the hard (big dot) and soft (small dot) particles, the black circle
  is the hard stable cone. The final hard jet is represented by the
  shaded area. The left (a) (centre (b), right (c)) plot corresponds
  to $\Delta_{12}<R$ ($R<\Delta_{12}<\sqrt{2}R$,
  $\sqrt{2}R<\Delta_{12}<2R$).}
\label{fig:active_2point_cone_pict}
\end{figure}

As in section \ref{sec:active_1point_cone}, the split--merge procedure
first deals with the pure-ghost protojets overlapping with the hard
stable cone. For $f>f_{\text{max}}$, this again only leads to
splittings. Depending of the value of $\Delta_{12}$, different
situations are found as shown on figure
\ref{fig:active_2point_cone_pict}. If the $y$-$\phi$ coordinates of
the particles are $p_1\equiv(\Delta,0)$ and $p_2\equiv (0,0)$, the
geometrical objects that are present are: the circle centred on $p_1$
with radius $R/2$; the tangents to this circle at $y$-$\phi$
coordinates $(\Delta_{12}/4,\pm\frac12\sqrt{R^2-(\Delta_{12}/2)^2})$;
and, for $\Delta_{12}<R$, the ellipse of eccentricity $\Delta_{12}/R$
whose foci are $p_1$ and $p_2$ (given by the equation, $\Delta_{a1} +
\Delta_{a2} = R$, where $a$ is a point on the ellipse). For
$\Delta_{12}<R$ the area of the jet is given by the union of the
ellipse, the circle and the regions between the ellipse, the circle and
each of the tangents (fig.~\ref{fig:active_2point_cone_pict}a). For $R
<\Delta_{12}<\sqrt2 R$ it is given by the circle plus the region
between the circle, the two tangents and the line equidistant from
$p_1$ and $p_2$ (fig.~\ref{fig:active_2point_cone_pict}b). For $\sqrt2
R <\Delta_{12}<2 R$ it is given by the circle plus the region between
the circle and two tangents, up to the intersection of the two
tangents (fig.~\ref{fig:active_2point_cone_pict}c). Finally, for
$\Delta\ge 2R$, the area is $\pi R^2/4$.

\begin{figure}
\centerline{\includegraphics{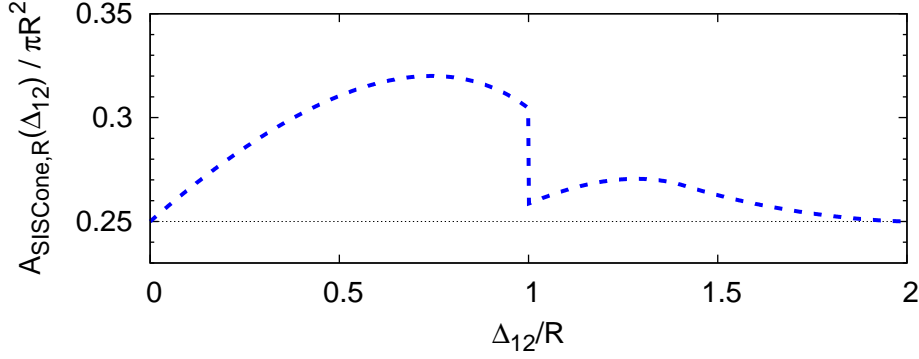}}
\caption{Active area of the hardest jet as a function of the distance
  between the hard and soft particle for the SISCone algorithm, \cf
  eq.~(\ref{eq:active_2points_cone}).}
\label{fig:active_2point_cone}
\end{figure}

An analytic computation of the active area gives
\begin{eqnarray}\label{eq:active_2points_cone}
  \frac{A_{\cone,R}(\Delta_{12})}{\pi R^2}
  & = & \frac{1}{4}
  \left[1-\frac{1}{\pi}\arccos\left(\frac{x}{2}\right)\right] \\
  & + & \begin{cases}
    \frac{x}{2\pi}\sqrt{1-\frac{x^2}{4}}
    +\frac{1}{4\pi}\sqrt{1-x^2}\arccos\left(\frac{x}{2-x^2}\right)
    & \Delta_{12}<R \\
    \frac{x}{2\pi}\sqrt{1-\frac{x^2}{4}}
    -\frac{x}{8\pi}\frac{1}{\sqrt{1-\frac{x^2}{4}}}
    & R<\Delta_{12}\le \sqrt{2}R \\
    \frac{1}{2\pi x}\sqrt{1-\frac{x^2}{4}}
    & \sqrt{2}R < \Delta_{12} \le 2R
  \end{cases}.\nonumber
\end{eqnarray}
with $x=\Delta_{12}/R$, while for $\Delta_{12}>2R$, one recovers the
result $\pi R^2/4$. The SISCone active area is plotted in
figure~\ref{fig:active_2point_cone} and it is compared to the results
for $k_t$ and Cambridge/Aachen in
figure~\ref{fig:2point-areas-active-3alg}. One notes that the SISCone
result is both qualitatively and quantitatively much further from the
passive result than was the case for $k_t$ and Cambridge/Aachen.

The main reason why the 2-point active area is larger than the 1-point
active area (whereas we saw the opposite behaviour for the passive
areas) is that the presence of the 2-particle configuration causes a
number of the pure-ghost cones that were present in the 1-particle case
to now contain the second particle and therefore be unstable. Since
these pure ghost cones are responsible for reducing the jet area
relative to the passive result (during the split--merge step), their
absence causes the active area to be `less-reduced' than in the
1-particle case.


\subsection{Area scaling violation}
\label{sec:area-scal-viol-active}

We can write the order $\as$ contribution to the average active area
in a manner similar to the passive area case,
eq.~(\ref{eq:ajfr}):
\begin{equation}
  \label{eq:Ajfr-base}
    \langle A_{\text{JA},R}\rangle =  
    A_{\text{JA},R}(0) + \langle \Delta
    A_{\text{JA},R}\rangle\,, 
\end{equation}
where we have used the relation
$A_{\text{JA},R}(\text{one-particle-jet})\equiv A_{\text{JA},R}(0)$
and where
\begin{equation}
  \label{eq:Ajfr}
  \langle \Delta  A_{\text{JA},R}\rangle \simeq \int_0 d\Delta_{12} 
  \int_{{Q_0}/\Delta_{12}}^{p_{t1}} d p_{t2} 
  \frac{dP}{dp_{t2} \, d\Delta_{12}} (
  A_{\text{JA},R}(\Delta_{12}) 
  - A_{\text{JA},R}(0))\,.
\end{equation}
Note that compared to eq.~(\ref{eq:ajfr}) we have removed the explicit
upper limit at $2R$ on the $\Delta_{12}$ integral since, for active
areas of sequential recombination algorithms,
$(A_{\text{JA},R}(\Delta_{12}) - A_{\text{JA},R}(0))$ may be 
non zero even for $\Delta_{12}>2R$.
Note also the notation for averages: we use $\avg{\cdots}$ to
refer to an average over perturbative emissions, while
$\avg{\cdots}_g$, implicitly contained in $A_{\JA,R}$ (see 
eq.~(\ref{eq:act_area})), refers to an average over ghost ensembles.
We now proceed as in section~\ref{sec:area-scal-viol-passive}, and write
\begin{equation}
  \label{eq:delta-Ajf-res}
  \langle \Delta A_{\text{JA},R}\rangle \simeq
  D_{\text{JA},R} \frac{C_1}{\pi b_0}
  \ln \frac{\as({Q_0})}{\as(R p_{t1})}\;, \qquad
  D_{\text{JA},R} = \int_0 \frac{d\theta}{\theta} ( 
  A_{\text{JA},R}(\theta)
  -  A_{\text{JA},R}(0))\,,
\end{equation}
where for brevity we have given just the running-coupling result.
One observes that the result continues to depend on $Q_0$, an
indication that the active area is an infrared-unsafe quantity, just
like the passive area.\footnote{We believe that any sensible (\ie
  related to the jet's sensitivity to UE/pileup type contamination)
  definition of area would actually be infrared unsafe.}
The coefficients for the anomalous dimension of the  active area for
the various algorithms are
\begin{subequations}
  \label{eq:Dalg1}
  \begin{align}
    D_{k_t,R} &\simeq 0.52 \, \pi R^2\,,
    \\    
    D_{\cam,R} &\simeq 0.08 \, \pi R^2\,, 
    \\ 
    D_{\cone,R} &\simeq 0.1246 \, \pi R^2 \,,
  \end{align}
\end{subequations}
where the SISCone result has been obtained by integrating the analytical
result, eq.~(\ref{eq:active_2points_cone}), while the results for
$k_t$ and Cambridge/Aachen have been  obtained both
by integrating the 2-point active-area results shown in
fig.~\ref{fig:2point-areas-active-3alg} and by a direct Monte Carlo
evaluation of eq.~(\ref{eq:delta-Ajf-res}). Note that while the
coefficients for $k_t$ and Cambridge/Aachen are only slightly different
from their passive counterparts, the one for SISCone has the opposite sign
relative to the passive one.

The treatment of higher-order fluctuations of active areas is more
complex than that for the passive ones, where the one-particle area
was a constant. We can separate the fluctuations of active areas into
two components, one (described above) that is the just the
one-particle result and the other, $\langle \Delta \Sigma^2_{\JA,R}
\rangle$, accounting for their modification in the presence of
perturbative radiation:
\begin{equation}
  \label{eq:act-fluct-seq}
  \langle \Sigma^2_{\JA,R} \rangle =
  \Sigma^2_{\JA,R}(0) + 
  \langle \Delta \Sigma^2_{\JA,R} \rangle \,,
\end{equation}
where $\Sigma_{\JA,R}(0)$ is given by eqs.~(\ref{eq:ktcamsigmas}a,b)
for the $k_t$ and the
Cambridge/Aachen algorithms respectively, 
and it is equal to zero for the SISCone algorithm.
The perturbative modification $\langle \Delta \Sigma^2_{\JA,R} \rangle$ 
is itself now driven by two mechanisms:
the fact that the second particle causes the average area to change,
and that it also causes modifications of the fluctuations associated
with the sampling over many ghost sets. We therefore write
\begin{align}
  \label{eq:sigma2_pt}
  \langle \Delta \Sigma^2_{\text{JA},R}\rangle &\simeq
  S_{\text{JA},R}^2 \frac{C_1}{\pi b_0}
  \ln \frac{\as({Q_0})}{\as(R p_{t1})}
  \;,\\ \qquad
  \label{eq:sigma2_pt_mid}
  S_{\text{JA},R}^2 &= \int_0 \frac{d\theta}{\theta} \big[
  (A_{\text{JA},R}(\theta) -  A_{\text{JA},R}(0))^2 + 
  \Sigma^2_{\text{JA},R}(\theta) -  \Sigma^2_{\text{JA},R}(0)
  \big]\, \\
  &= \int_0 \frac{d\theta}{\theta} 
  (A^2_{\text{JA},R}(\theta) -  A^2_{\text{JA},R}(0))
  -2 A_{\text{JA},R}(0) D_{\text{JA},R}\, ,
  \label{eq:sigma2_pt_end}
\end{align}
where as usual we neglect contributions that are not enhanced by any
logarithm or that are higher-order in $\as$. The details of how to
obtain these results are given in Appendix \ref{sec:app_fluct}.

The coefficient $S_{\text{JA},R}^2$ can  be determined only numerically for the $k_t$ and
Cambridge/Aachen algorithms, while for the SISCone algorithm the result
can be deduced from eq.~(\ref{eq:active_2points_cone}) together with
the knowledge that $\Sigma_{\cone,R}(\theta) \equiv 0$:
\begin{subequations}
  \label{eq:Sigmaalg1}
  \begin{align}
    {S}_{\kt,R}^2 &\simeq (0.41 \, \pi R^2)^2\,,\\ 
    {S}_{\cam,R}^2 &\simeq  (0.19 \, \pi R^2)^2\,,\\ 
    {S}_{\cone,R}^2 &\simeq  (0.0738\, \pi R^2)^2\,.
  \end{align}
\end{subequations}
Again, both the values and their ordering are similar to what we have
obtained for the passive areas (see eq.~\eqref{eq:fluct_passive_coefs}).

\subsection{$\bs n$-particle properties and  large-$\bs n$ behaviour}
\label{sec:n-particle-active}

\subsubsection{$\bs{k_t}$ algorithm}
\label{sec:kt-large-n-active}

As for the passive area, the $k_t$ algorithm's active area has the
property that it can be expressed as a sum of individual particle
areas:
\begin{equation}
  \label{eq:particle-areas-active}
  A_{\kt,R}(J) = \sum_{p_i \in J} A_{\kt,R}(p_i)\,.
\end{equation}
This is the case because the presence of the momentum scale $k_t$ in
the distance measure means that all ghosts cluster among themselves
and with the hard particles, before any of the hard particles start
clustering between themselves. However in contrast to the passive-area
situation, there is no known simple geometrical construction for the
individual particle area.

\subsubsection{Equivalence of all areas for large $\bs{n}$}
\label{sec:large-n-equiv}

The existence of different area definitions is linked to the ambiguity
in assigning `empty space' to any particular jet. In the presence of a
sufficiently large number of particles $n$, one expects this ambiguity to
vanish because real particles fill up more and more of the empty space
and thus help to sharpen the boundaries between jets. Thus in the
limit of many particles, all sensible area definitions should give
identical results for a given jet.

To quantify this statement, we examine (a bound on) the
scaling law for the relation between the density of particles and the
magnitude of the potential differences between different area
definitions.
We consider the limit of `dense' coverage, defined as follows: take
square tiles of some size and use them to cover the rapidity--azimuth
cylinder (up to some maximal rapidity).  Define $\lambda$ as the
smallest value of tile edge-length such that all tiles contain at
least one particle. In an event consisting of uniformly distributed
particles, $\lambda$ is of the same order of magnitude as the typical
interparticle distance. The event is considered to be dense if
$\lambda \ll R$.

Now let us define a boundary tile of a jet to be a tile that contains
at least one particle of that jet and also contains a particle from
another jet or has an adjacent tile containing one or more
particle(s) belonging to a different jet.
We expect that the difference between different jet-area definitions
cannot be significantly larger than
the total area of the boundary tiles for a jet. 

The number of boundary tiles for jets produced by a given jet
algorithm (of radius $R$) may scale in a non-trivial manner with the
inter-particle spacing $\lambda$, since the boundary may well have a
fractal structure. We therefore parametrise the average number of
boundary tiles for a jet, $N_{b,\JA,R}$ as
\begin{equation}
  \label{eq:Nb}
  N_{b,\JA,R} \sim \left(\frac{R}{\lambda}\right)^{{\!\daleth}_{\,\JA}},
\end{equation}
where the fractal dimension ${\!\daleth}_{\,\JA}=1$ would correspond to a smooth
boundary. The total area of these boundary tiles gives an upper limit
on the ambiguity of the jet area
\begin{equation}
  \label{eq:area-diff-upper-bound}
   \langle |a_{\JA,R} - A_{\JA,R}| \rangle \lesssim N_{b,\JA,R}\, \lambda^2 \sim 
      R^{{\!\daleth}_{\,\JA}} \lambda^{2-{\!\daleth}_{\,\JA}}\,,
\end{equation}
and similarly for the difference between active or passive and Voronoi
areas.
As long as ${\!\daleth}_{\,\JA} < 2$ the differences between various
area definitions are guaranteed to vanish in the infinitely dense
limit, $\lambda \to 0$.  We note that ${\!\daleth}_{\,\JA} = 2$
corresponds to a situation in which the boundary itself behaves like
an area, \ie occupies the same order of magnitude of space as the jet
itself. This would be visible in plots representing jet active areas
(such as fig.~\ref{fig:example-active-area}), in the form of finely
intertwined jets. We have seen no evidence for this and therefore
believe that $1 \le {\daleth}_{\,\JA} < 2$ for all three jet
algorithms considered here.

In practice we expect the difference between any two area definitions
to vanish much more rapidly than eq.~(\ref{eq:area-diff-upper-bound}) as $\lambda \to 0$, since the
upper bound will only be saturated if, for every tile, the difference
between two area definitions has the same sign. This seems highly
unlikely. If instead differences in the area for each tile are
uncorrelated (but each of order $\lambda^2$) then one would expect to
see
\begin{equation}
  \label{eq:area-diff-guessed-bound}
   \langle |a_{\JA,R} - A_{\JA,R}| \rangle \sim \sqrt{N_{b,\JA,R}}\, \lambda^2 \sim 
      R^{{\!\daleth}_{\,\JA}/2} \lambda^{2-{\!\daleth}_{\,\JA}/2}\,.
\end{equation}
We have measured the fractal dimension for the $k_t$ and
Cambridge/Aachen algorithms and find ${\daleth}_{\,\kt} \simeq
{\!\daleth}_{\,\cam} \simeq 1.20-1.25$.\footnote{This has been
  measured on pure ghost jets, because their higher multiplicity
  facilitates the extraction of a reliable result, however we strongly
  suspect that it holds also for single-particle jets.}
Note that any measurement of the fractal dimension of jet algorithms
in real data would be severely complicated by additional structure
added to jets by QCD branching, itself also approximately fractal in
nature.

The fact that active and passive (or Voronoi) areas all give the same result
in dense events has practical applications in real-life situations where
an event is populated by a very large number of particles (heavy ion 
collisions being an example). In this case it will be possible to choose
the area type which is fastest to compute (for instance the Voronoi area) 
and  use the results in place of the active or passive one.

\section{Back reaction}
\label{sec:back-reaction}

So far we have considered how a set of infinitely soft ghosts clusters
with a hard jet, examining also cases where the jet has some
finitely-soft substructure. This infinitely-soft approximation for the
ghosts is not only adequate, but also necessary from the point of view
of properly defining jet areas.  
However if we are to understand the impact of minimum-bias (MB) and
underlying-event radiation on jet finding --- the original motivation
for studying areas --- then we should take into account the fact that
these contributions provide a dense background of particles at some
small but \emph{finite} soft scale $\sim \Lambda_{QCD}$. 

This has two main consequences. The first (more trivial) one is that
the minimum-bias or underlying event can provide an alternative,
dynamic infrared cutoff in the $p_{t2}$ integration in equations such
as~eq.~(\ref{eq:ajfr}): assuming that the density of MB transverse
momentum per unit area is given by $\rho$ then one can expect that
when $p_{t2} \ll \pi R^2 \rho$, the presence of $p_{2}$ will no
longer affect the clustering of the ghosts (\ie the MB particles). In
this case the $p_{t2}$ integral will then acquire an effective
infrared cutoff $\sim \pi R^2 \rho$ and in expressions such as
eqs.~(\ref{eq:delta-ajf-res},\ref{eq:delta-ajf-res-running}), $Q_0$
will be replaced by $\pi R^3 \rho$. 
Note that neither with infinitely nor finitely-soft ghosts do we
claim control over the coefficient in front of this cutoff, though we
do have confidence in the prediction of its $R$ dependence for small
$R$.\footnote{The
  coefficient can actually be determined quite straightforwardly,
  however since its impact is of the same order as other effects that
  we neglect (\ie free of any logarithmic enhancements) we leave its
  determination to future work.}

The second consequence of the finite softness of the MB contribution
is that the addition of the MB particles can modify the set of non-MB
particles that make it into a jet. We call this `back-reaction' and it
is the main subject of this section.

That back reaction should happen is quite intuitive as
concerns non-MB particles whose softness is commensurate with the MB
scale. However the extent to which it occurs depends significantly on
the jet algorithm. Furthermore for some algorithms it can also occur
(rarely) even for non-MB particles that are much harder than the MB
scale.

As with the studies of areas, there are two limits that can usefully be
examined for back-reaction: the illustrative and mathematically
simpler (but less physical) case in which the MB is pointlike and the
more realistic case with diffuse MB radiation.

\subsection{Back reaction from pointlike minimum-bias}
\label{sec:pointlike-back-reaction}

Let us first calculate back reaction in the case of pointlike minimum
bias. We will consider minimum-bias particles with transverse momentum
$p_{tm}$ distributed uniformly on the $y$--$\phi$ cylinder with density $\nu_m \ll
1$. We use a subscript $m$ rather than $g$ to differentiate them from
ghost particles, the key distinction being that $p_{tm}$ is small but
finite, where $p_{tg}$ is infinitesimal.

Let us consider the situation in which a particle $p_1$, with large
transverse momentum, $p_{t1} \gg p_{tm}$, has emitted a soft particle
$p_2$ on a scale commensurate with the minimum-bias particles, $p_{t2} \sim
p_{tm}$. We shall calculate the probability that $p_2$ was part of the
jet in the absence of the minimum-bias particle, but is \emph{lost}
from it when the minimum-bias particle is added. This can be written
\begin{equation}
  \label{eq:loss-passive-master}
  \frac{dP^{(L)}_{\JA,R}}{d p_{t2}} = 
  \int d\phi_m dy_m \nu_m \int d\Delta_{12} \frac{dP}{dp_{t2} \,
    d\Delta_{12}} H_{\JA,R}(p_2 \in J_1) \, 
  H_{\JA,R}(p_2 \notin J_1 | \,p_m)\,,
\end{equation}
where $H_{\JA,R}(p_2 \in J_1)$ is $1$ ($0$) if, in the absence of
$p_m$, $p_{2}$ is inside (outside) the jet that contains $p_1$.
Similarly, $H_{\JA,R}(p_2 \notin J_1 | p_m)$ is $1$ ($0$) if, in the
presence of $p_m$, $p_{2}$ is inside (outside) the jet that contains
$p_1$. One can also define the probability for $p_2$ to not be part of
the jet in the absence of the minimum-bias particle, but to be
\emph{gained} by the jet when the minimum-bias particle is added,
\begin{equation}
  \label{eq:gain-passive-master}
  \frac{dP^{(G)}_{\JA,R}}{d p_{t2}} = 
  \int d\phi_m dy_m \nu_m \int d\Delta_{12} \frac{dP}{dp_{t2} \,
    d\Delta_{12}} H_{\JA,R}(p_2 \notin J_1) \, 
  H_{\JA,R}(p_2 \in J_1 | \,p_m)\,.
\end{equation}
It is convenient to factor out the particle production probability as
follows, in the small $R$ limit,
\begin{equation}
  \label{eq:loss-passive-factorize}
  \frac{dP^{(L)}_{\JA,R}}{d p_{t2}} = \left. \Delta_{12} \frac{dP}{dp_{t2}
      \, d\Delta_{12}}
  \right|_{\Delta_{12}=R} \nu_m \; b^{(L)}_{\JA,R}(p_{t2}/p_{tm})\,,
\end{equation}
where $b^{(L)}_{\JA,R}(p_{t2}/p_{tm})$ can be thought of as the
effective `back-reaction area' over which the minimum-bias particle
causes a loss of jet contents, given a $d\Delta_{12}/\Delta_{12}$
angular distribution for the jet contents:\footnote{Strictly speaking
  it is the integral over area of the probability of causing a loss of
  jet contents.}
\begin{equation}
  \label{eq:loss-effective-area}
  b^{(L)}_{\JA,R}(p_{t2}/p_{tm}) = \int d\phi_m dy_m \int
  \frac{d\Delta_{12}}{\Delta_{12}} 
  H_{\JA,R}(p_2 \in J_1) \, 
  H_{\JA,R}(p_2 \notin J_1 | \,p_m)\,.
\end{equation}
One can similarly define an effective back-reaction area for gain,
\begin{align}
  \frac{dP^{(G)}_{\JA,R}}{d p_{t2}} &= \left. \Delta_{12} \frac{dP}{dp_{t2}
      \, d\Delta_{12}}
  \right|_{\Delta_{12}=R} \nu_m \; b^{(G)}_{\JA,R}(p_{t2}/p_{tm})\,,
  \\[4pt]
  \label{eq:gain-effective-area}
  b^{(G)}_{\JA,R}(p_{t2}/p_{tm}) &= \int d\phi_m dy_m \int
  \frac{d\Delta_{12}}{\Delta_{12}} 
  H_{\JA,R}(p_2 \notin J_1) \, 
  H_{\JA,R}(p_2 \in J_1 | \,p_m)\,.  
\end{align}

For sequential-recombination algorithms the $H$ functions in
eqs.~(\ref{eq:loss-effective-area},\ref{eq:gain-effective-area})
translate to a series of $\Theta$-functions, \eg
\begin{multline}
  \label{eq:HL-kt}
  H_{\kt,R}(p_2 \in J_1) H_{\kt,R}(p_2 \notin J_1 |\, p_m) 
    = \Theta(R - \Delta_{12}) \Theta(\Delta_{1(2+m)} - R) \times \\
    \times
    \Theta(\Delta_{1m}- \min(1,p_{t2}/p_{tm}) \Delta_{2m}) 
    \Theta(\Delta_{12}- \min(1,p_{tm}/p_{t2}) \Delta_{2m}) 
    \Theta(R-\Delta_{2m})\,,
\end{multline}
for the $k_t$ algorithm and 
\begin{multline}
  \label{eq:HL-cam}
  H_{\cam,R}(p_2 \in J_1) H_{\cam,R}(p_2 \notin J_1 |\, p_m) 
    = \Theta(R - \Delta_{12}) \Theta(\Delta_{1(2+m)} - R) \times \\
    \times
    \Theta(\Delta_{1m}-  \Delta_{2m}) 
    \Theta(\Delta_{12}-  \Delta_{2m}) 
    \Theta(R-\Delta_{2m})\,,
\end{multline}
for Cambridge/Aachen, where $\Delta_{1(2+m)}$ is the distance between
$p_1$ and the recombined $p_{2}+p_{m}$. Evaluating integrals with the
above $\Theta$-functions is rather tedious,
but one can usefully consider the limit $p_{tm} \ll
p_{t2} \ll p_{t1}$. This of physical interest because it relates to
the probability that the minimum-bias particle induces changes in jet momentum
that are much larger than $p_{tm}$, and a number of simplifications
occur in this limit. Since
\begin{equation}
  \label{eq:BR-distance-approx-large-pt2}
  \Delta_{1(2+m)} = 
  \left|\vec \Delta_{12} + \frac{p_{tm}}{p_{t2}}\vec \Delta_{2m}\right|
  = \Delta_{12} + \frac{p_{tm}}{p_{t2}} \frac{\vec \Delta_{12} \cdot \vec
  \Delta_{2m}}{\Delta_{12}} + \order{\frac{p_{tm}^2}{p_{t2}^2} R}\,,
\end{equation}
$p_2$ must be close to the edge of the jet in order for it to be
pulled out by $p_m$, $|\Delta_{12}-R| \ll 1$. Without loss of
generality, we can set 
$y_1=\phi_1=\phi_2=0$, so that $\Delta_{12} = y_2 \simeq R$, and
\begin{equation}
  \Delta_{1(2+m)}
  = y_2 + \frac{p_{tm}}{p_{t2}} (y_m - R) +
  \order{\frac{p_{tm}^2}{p_{t2}^2} R}.
\end{equation}
We can then carry out the integrations over $\phi_m$ and $y_2$
straightforwardly, leading to following the result for loss at high
$p_{t2}$,
\begin{equation}
  \label{eq:bLres}
  b^{(L)}_{\kt,R}(p_{t2}/p_{tm} \gg 1) \simeq 
  b^{(L)}_{\cam,R}(p_{t2}/p_{tm} \gg 1) \simeq
   \int_{R}^{2R} d y_m \, \frac{(y_m-R)}{R} \frac{p_{tm}}{p_{t2}}\, 
   2\sqrt{R^2-(y_m-R)^2} =
    \frac{2}{3} \frac{p_{tm}}{p_{t2}} R^2\,.
\end{equation}
In a similar manner one obtains for the gain,
\begin{multline}
  \label{eq:bGres}
  b^{(G)}_{\kt,R}(p_{t2}/p_{tm} \gg 1) \simeq 
  b^{(G)}_{\cam,R}(p_{t2}/p_{tm} \gg 1) \simeq
   \int_{R/2}^{R} d y_m \, \frac{(R-y_m)}{R} \frac{p_{tm}}{p_{t2}} \, 
   2\sqrt{R^2-(R-y_m)^2} =\\=
    \left(\frac{2}{3} - \frac{\sqrt{3}}{4}\right) \frac{p_{tm}}{p_{t2}} R^2\,.
\end{multline}
\begin{figure}
  \centering
  \includegraphics[width=0.48\textwidth]{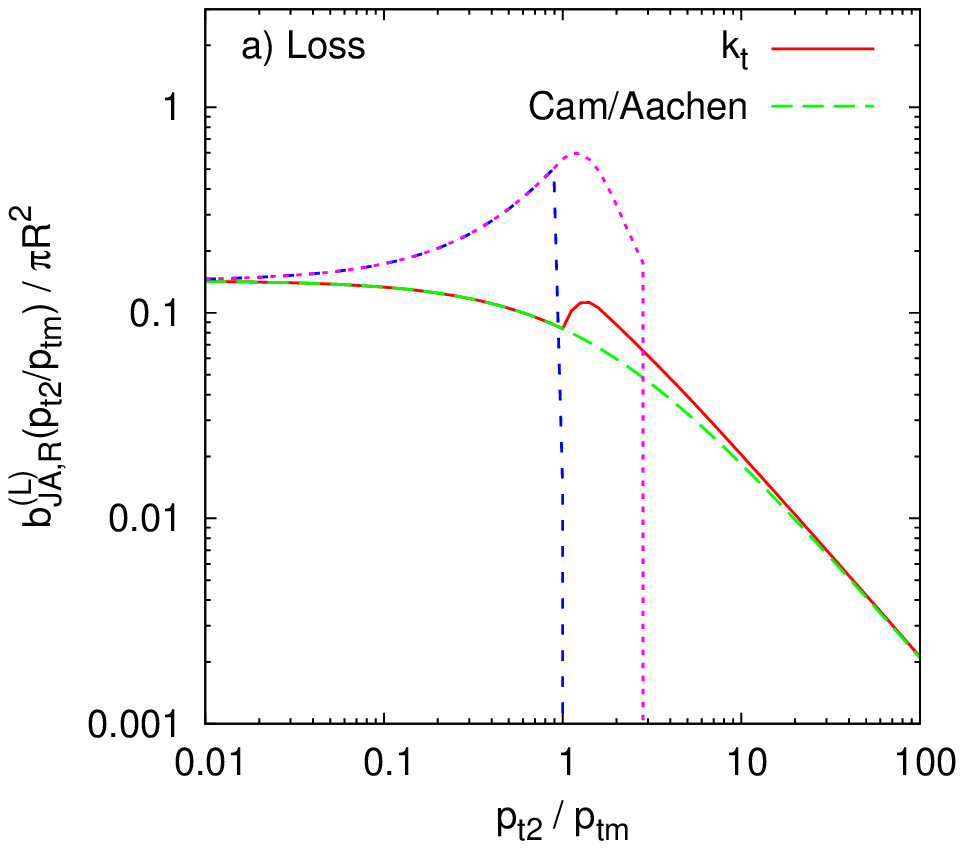}
  \hfill
  \includegraphics[width=0.48\textwidth]{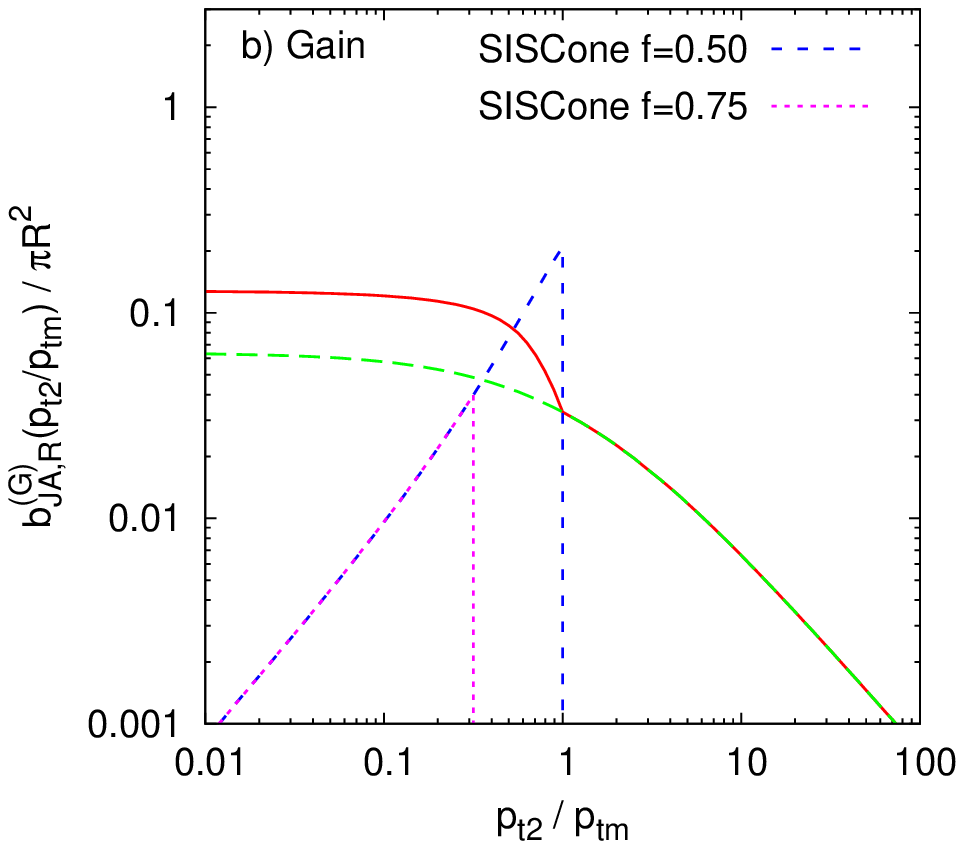}
  \caption{The effective area for back-reaction as a function of the
    ratio of the soft perturbative scale $p_{t2}$ and the point-like
    minimum-bias scale $p_{tm}$, showing separately the loss (a) and
    gain (b) components for four jet definitions.}
  \label{fig:passive-loss-gain}
\end{figure}%
The results for general $p_{t2}/p_{tm}$, determined numerically, are
shown in figure~\ref{fig:passive-loss-gain}. The SISCone results are
included and have the property that\,,
\begin{subequations}
\label{eq:cone-GL-cutoff}
  \begin{align}
    b^{(L)}_{\cone,R}(p_{t2}/p_{tm}) &= 0 \quad \text{for} \quad
    \frac{p_{t2}}{p_{tm}} > \frac{f}{1-f}\,,\\
    b^{(G)}_{\cone,R}(p_{t2}/p_{tm}) &= 0 \quad \text{for} \quad
    \frac{p_{t2}}{p_{tm}} > \frac{1-f}{f}\,.
  \end{align}
\end{subequations}
\ie at high $p_{t2}$, point-like minimum bias never induces
back-reaction in the cone algorithm,\footnote{For large but finite
  $p_{t1}$, back-reaction can actually occur beyond the above limits,
  but only with probability $\sim p_{tm}/p_{t1}$.} in contrast to the
situation with the sequential recombination algorithms, for which
back-reaction occurs with a suppressed, but non-zero probability $\sim
p_{tm}/p_{t2}$. On the other hand, for $p_{t2} \sim p_{tm}$
back-reaction is more likely with the cone algorithm --- the effective
area over which the MB particle can cause a change in jet contents is $\sim
0.5 \pi R^2$, to be compared to $\sim 0.1 \pi R^2$ for the $\kt$ and
Cambridge/Aachen algorithms.

One may use the results
eqs.~(\ref{eq:bLres})--(\ref{eq:cone-GL-cutoff}) to determine the
average change in jet-momentum due to back reaction.
Because of the logarithmic spectrum of emissions $dP/(dp_{t2} \,
d\Delta_{12})$, one finds that it receives contributions from the whole
logarithmic region $p_{tm} < p_{t2} < p_{t1}$,
\begin{equation}
  \label{eq:deltapt_back}
  \langle \Delta p_{t,\JA,R}^{(G-L)} \rangle \simeq
  \int_{p_{tm}}^{p_{t1}} dp_{t2} p_{t2} \left[\frac{dP^{(G)}_{\JA,R}}{dp_{t2}} -
    \frac{dP^{(L)}_{\JA,R}}{dp_{t2}} \right] =
  \beta_{\JA,R} \, 
  \rho  \cdot \frac{C_1}{\pi b_0} \ln
  \frac{\as(p_{tm} R)}{\as(p_{t1} R)}\,,
\end{equation}
(evaluated for fixed coupling), where $\rho = \nu_m p_{tm}$
corresponds to the average transverse momentum of minimum-bias
radiation per unit area and
\begin{equation}
  \label{eq:passive-beta}
  \beta_{\JA,R} = \lim_{p_{t2} \to \infty} \frac{p_{t2}}{p_{tm}}
  \left(b^{(G)}_{\JA,R}(p_{t2}/p_{tm}) -
    b^{(L)}_{\JA,R}(p_{t2}/p_{tm}) \right)\,. 
\end{equation}
The structure of the correction in eq.~(\ref{eq:deltapt_back}) is very
similar to that for the actual contamination from minimum bias, $\rho
\langle \Delta a_{\JA,R}\rangle$, with $ \langle \Delta
a_{\JA,R}\rangle$ as determined in
section~\ref{sec:area-scal-viol-passive}: notably, for fixed coupling,
the average back-reaction
scales with the logarithm of the jet $p_t$.  The coefficients
$\beta_{\JA,R}$,
\begin{subequations}
  \label{eq:passive-beta-res}
  \begin{align}
    \beta_{\kt,R} = \beta_{\cam,R} &= -\frac{\sqrt{3}}{4} R^2 \simeq
    -0.1378 \pi R^2\,,\\
    \beta_{\cone,R} &= 0\,.
  \end{align}
\end{subequations}
can be directly compared to the results for the $d_{\JA,R}$ there. The
values are relatively small, similar in particular to what one observes
for the Cambridge/Aachen algorithm, though of opposite sign.

Though the average change in jet momentum,
both from scaling violations of the area and from back-reaction, have
a similar analytical structure,
it is worth bearing in mind that these similar analytical structures
come about quite differently in the two cases.
Regarding area scaling violations, a significant fraction of jets,
$\sim \as \ln p_{t1}/p_{tm}$, are subject to a change in area $\sim
R^2$ (\cf section~\ref{sec:area-scal-viol-passive}), and a consequent
modification of the minimum-bias contamination by a modest amount
$\sim p_{tm}$.
In contrast the average back reaction effect $\sim \as p_{tm} \ln
p_{t1}/p_{tm}$, is due to large modifications of the jet momentum
$\sim p_{t2}$ with $p_{tm} \ll p_{t2} \ll p_{t1}$ occurring rarely,
with a differential probability that depends on $p_{t2}$ as $\sim \as
d p_{t2} p_{tm}/p_{t2}^2$.
One consequence of this is that the mean square change in transverse
momentum due to back-reaction is dominated by very rare modifications
of jets in which $p_{t2} \sim p_{t1}$, giving
\begin{equation}
  \label{eq:BR-dispersion}
  \left\langle \left(\Delta p_{t,\JA,R}^{(G,L)}\right)^2
  \right\rangle \sim \as p_{t1} p_{tm} \nu_m\,.
\end{equation}
Note that the coefficient of this dispersion is non-zero even for the
SISCone algorithm, due to the residual probability $\sim p_{tm}/p_{t1}$
that it has, for finite $p_{t1}$, for any change in structure with
$p_{tm} \ll p_{t2} \lesssim p_{t1}$.

\subsection{Back reaction from diffuse MB}
\label{sec:diffuse-back-reaction}

Let us now examine back reaction in the case where the minimum-bias
radiation has a uniform diffuse structure, consisting of a high
density of minimum-bias particles, $\nu_m \gg 1$. The relation between
back reaction in the point-like and diffuse cases is rather similar to
the relation between passive and active areas: key analytical features
remain unaffected by the change in the structure of the minimum bias,
but certain specific coefficients change.

We define the probability for loss in the presence of diffuse MB as
\begin{equation}
  \label{eq:BR-active-loss-master}
  \frac{dP^{(L)}_{\JA,R}}{d p_{t2}} = 
  \mathop{\lim_{\nu_m \to \infty}}_{\rho = \nu_m \langle
    p_{tm}\rangle\: \text{fixed}}
  \left\langle 
    \int d\Delta_{12} \frac{dP}{dp_{t2} \,
      d\Delta_{12}} H_{\JA,R}(p_2 \in J_1) \, 
    H_{\JA,R}(p_2 \notin J_1 | \,\rho)
  \right\rangle_{\mathrm{MB}},
\end{equation}
where the average is performed over the ensemble of MB configurations,
and now $H_{\JA,R}(p_2 \notin J_1 | \,\rho)$ is 1 (0) if $p_2$ is
outside (inside) the jet containing $p_1$ in the presence of the
specific minimum-bias instance. A similar equation holds for the gain
probability.

Then, as with eq.~(\ref{eq:loss-passive-factorize}) we factorise this,
\begin{equation}
  \label{eq:loss-active-factorize}
  \frac{dP^{(L)}_{\JA,R}}{d p_{t2}} = \left. \Delta_{12} \frac{dP}{dp_{t2}
      \, d\Delta_{12}}
  \right|_{\Delta_{12}=R} \; B^{(L)}_{\JA,R}(p_{t2} / \rho)\,,
\end{equation}
into one piece related to the probability for perturbative emission
and a second piece $B^{(L)}_{\JA,R}$ that is the diffuse analogue of
the effective back-reaction area $b^{(L)}_{\JA,R}$ in the point-like
case.  Note however that it is not so obvious exactly what geometrical
area $B^{(L)}_{\JA,R}$ actually corresponds to.\footnote{A related
  issue is that the precise choice of normalisation of
  $B^{(L)}_{\JA,R}$ is somewhat arbitrary --- our specific choice is
  intended to provide a meaningful connection with $b^{(L)}_{\JA,R}$
  in the large $p_{t2}$ limit.} 
Similarly, we introduce $dP^{(G)}_{\JA,R}/d p_{t2}$ and
$B^{(G)}_{JA,R}$ corresponding to the gain in the presence of a
diffuse MB.

The only case for which we have analytical results for
$B^{(L,G)}_{\JA,R}(p_{t2} / \rho)$ is the SISCone algorithm, for which
(with $f > f_{\max}\simeq 0.391$)
\begin{equation}
  \label{eq:cone-active-BR-is-zero}
  B^{(L)}_{\cone,R}(p_{t2} / \rho) = B^{(G)}_{\cone,R}(p_{t2} / \rho)
  = 0\,.
\end{equation}
This is a consequence of the facts (a) that the addition of a uniform
background of MB particles has no effect on the stability (or
instability) of a specific cone, (b) that for $p_{t2} \ll p_{t1}$
the split--merge step is immaterial to the jet finding if $p_2$ is within the
cone around $p_1$, and (c) that for $p_{2}$ outside the cone around
$p_1$, the maximal possible overlap is of $p_{2}$'s stable cone with
that of $p_{1}$ is $f_{\max}$ and if $f>f_{\max}$ then the two cones
will always be split, ensuring that $p_{2}$ remains in a jet distinct
from $p_{1}$. We believe that real-life corrections to the zero in
eq.~(\ref{eq:cone-active-BR-is-zero}) are proportional to the standard
deviation of MB transverse-momentum density from point to point in the
event.

\begin{figure}
  \centering
  \includegraphics[width=0.48\textwidth]{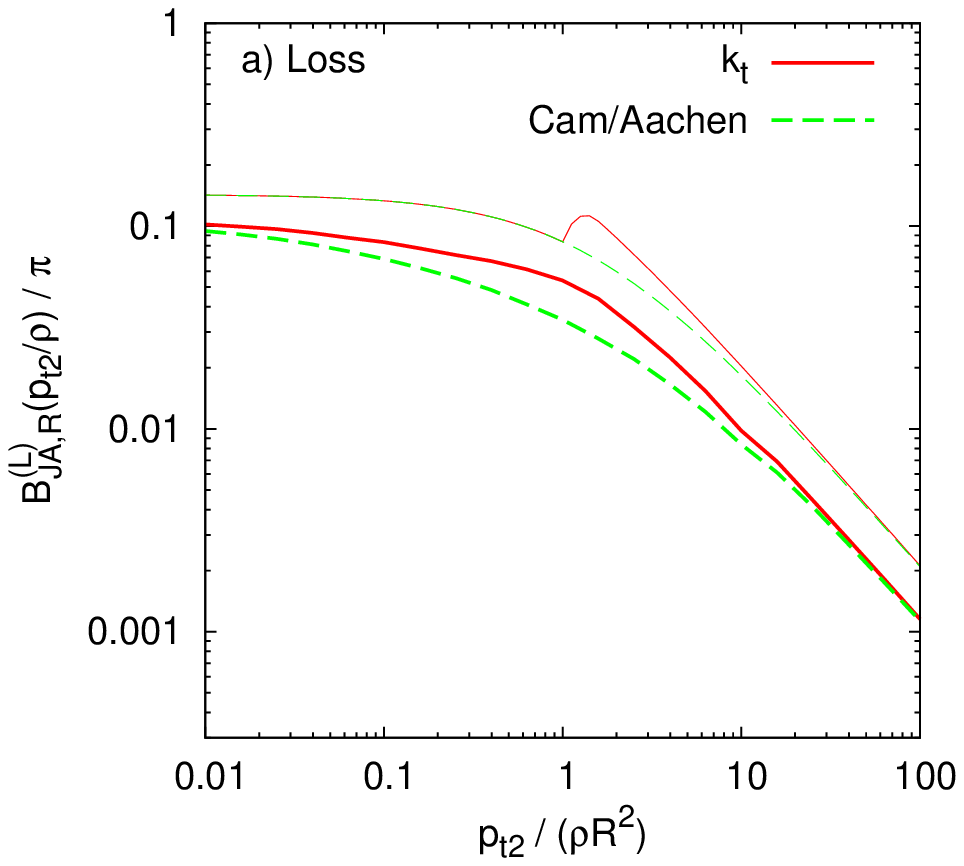}
  \hfill
  \includegraphics[width=0.48\textwidth]{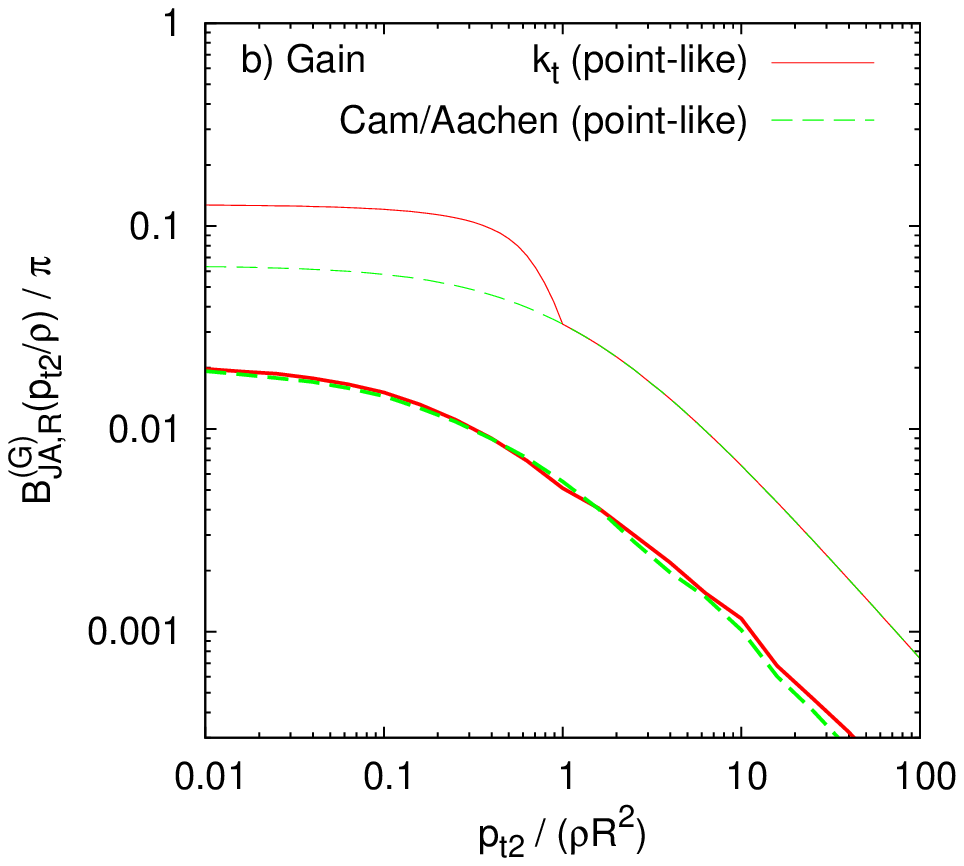}
  \caption{Numerical results for the diffuse effective back-reaction
    `area', $B^{(L,G)}_{\JA,R} (p_{t2}/\rho)$, for the $k_t$ and
    Cambridge/Aachen algorithms, with the point-like results
    $b^{(L,G)}_{\JA,1} (p_{t2}/p_{tm})$ shown for comparison
    also. Results obtained for $R=1$ and verified also for $R=0.7$.}
  \label{fig:active-loss-gain}
\end{figure}%

Numerical results for the $B^{(L,G)}_{\JA,R}(p_{t2} / \rho)$ for the
$k_t$ and Cambridge/Aachen algorithms are given in in
figure~\ref{fig:active-loss-gain} and compared to the results in the
point-like case. One sees that the general functional form is rather
similar though the normalisations are somewhat smaller (by a factor of
$2$ for loss, a factor $\sim 10$ for gain). The asymptotic
large-$p_{t2}$ behaviours are observed to be
\begin{subequations}
  \label{eq:BLG-asymptotic}
  \begin{align}
    B^{(L)}_{\kt,R}(p_{t2} / \rho) \simeq B^{(L)}_{\cam,R}(p_{t2} / \rho) &\simeq 0.11 \,\pi R^2 \, \frac{\rho}{p_{t2}}\\
    B^{(G)}_{\kt,R}(p_{t2} / \rho) \simeq B^{(G)}_{\cam,R}(p_{t2} / \rho) &\simeq 0.013 \,\pi R^2 \,\frac{\rho}{p_{t2}}
  \end{align}
\end{subequations}
As in the point-like case we can calculate the mean change in jet
transverse momentum due to back reaction and we obtain
\begin{equation}
  \label{eq:deltapt_back_active}
  \langle \Delta p_{t,\JA,R}^{(G-L)} \rangle \simeq
  \int_{p_{tm}}^{p_{t1}} dp_{t2} p_{t2} \left[\frac{dP^{(G)}_{\JA,R}}{dp_{t2}} -
    \frac{dP^{(L)}_{\JA,R}}{dp_{t2}} \right] =
  {\mathcal{B}}_{\JA,R} \, 
  \rho  \cdot \frac{C_1}{\pi b_0} \ln
  \frac{\as(\rho R^3)}{\as(p_{t1} R)}\,,
\end{equation}
with 
\begin{equation}
  \label{eq:active-beta}
  {\mathcal{B}}_{\JA,R} = \lim_{p_{t2} \to \infty} \frac{p_{t2}}{\rho}
  \left(B^{(G)}_{\JA,R}(p_{t2}/\rho) -
    B^{(L)}_{\JA,R}(p_{t2}/\rho) \right)\,. 
\end{equation}
Even though $b^{(L,G)}_{\JA,R}$ had $p_{t2}/p_{tm}$ as its argument
and $B^{(L,G)}_{\JA,R}$ has $p_{t2}/\rho$, the final expressions for
the average back-reaction in the point-like and diffuse cases,
eqs.~(\ref{eq:deltapt_back}), (\ref{eq:deltapt_back_active}), have
almost identical forms --- in particular the overall scale appearing
in each is $\rho$ and the only difference appears in the denominator
for the argument of the logarithm. The coefficients are slightly smaller,
\begin{subequations}
  \label{eq:active-beta-res}
  \begin{align}
    {\mathcal{B}}_{\kt,R} = {\mathcal{B}}_{\cam,R} &\simeq
    -0.10 \pi R^2\,,\\
    {\mathcal{B}}_{\cone,R} &= 0\,,
  \end{align}
\end{subequations}
and will again translate to modest effects compared to the overall
minimum-bias contamination in the jets.
Note also however that in general the scaling with $R$ of
$b^{(L/G)}_{\kt,R}(p_{t2} / p_{tm})$ and $B^{(L/G)}_{\kt,R}(p_{t2} /
\rho)$ is subtly different. The former truly behaves like an area, in
that $b^{(L/G)}_{\kt,R}(p_{t2} / p_{tm})/ R^2$ is $R$-independent; the
latter instead has the property that it is $B^{(L/G)}_{\kt,R}(R^2
p_{t2} / \rho)$ that is $R$-independent. 

\section{Areas in (simulated) real life}
\label{sec:real-life}

In this section we shall examine the properties of jet areas in the
context of realistic events, as simulated with Herwig \cite{Herwig}
and Pythia \cite{Pythia}. There are two purposes to doing so. 
Firstly we wish to illustrate the extent to which the
simple single-gluon emission arguments of the previous section hold
once one includes full parton-showering and hadronisation.
Secondly, jet areas have the potential to play an important role in
the estimation and subtraction of underlying event and pileup
contamination, as discussed in \cite{AreaSubtraction}. The study of
jet areas in realistic events can help to highlight some of the issues
that arise in such a procedure.

\subsection{Jet area distributions and anomalous dimension}

\begin{figure}[tp]
  \centering
  \includegraphics[height=\textwidth,angle=270]{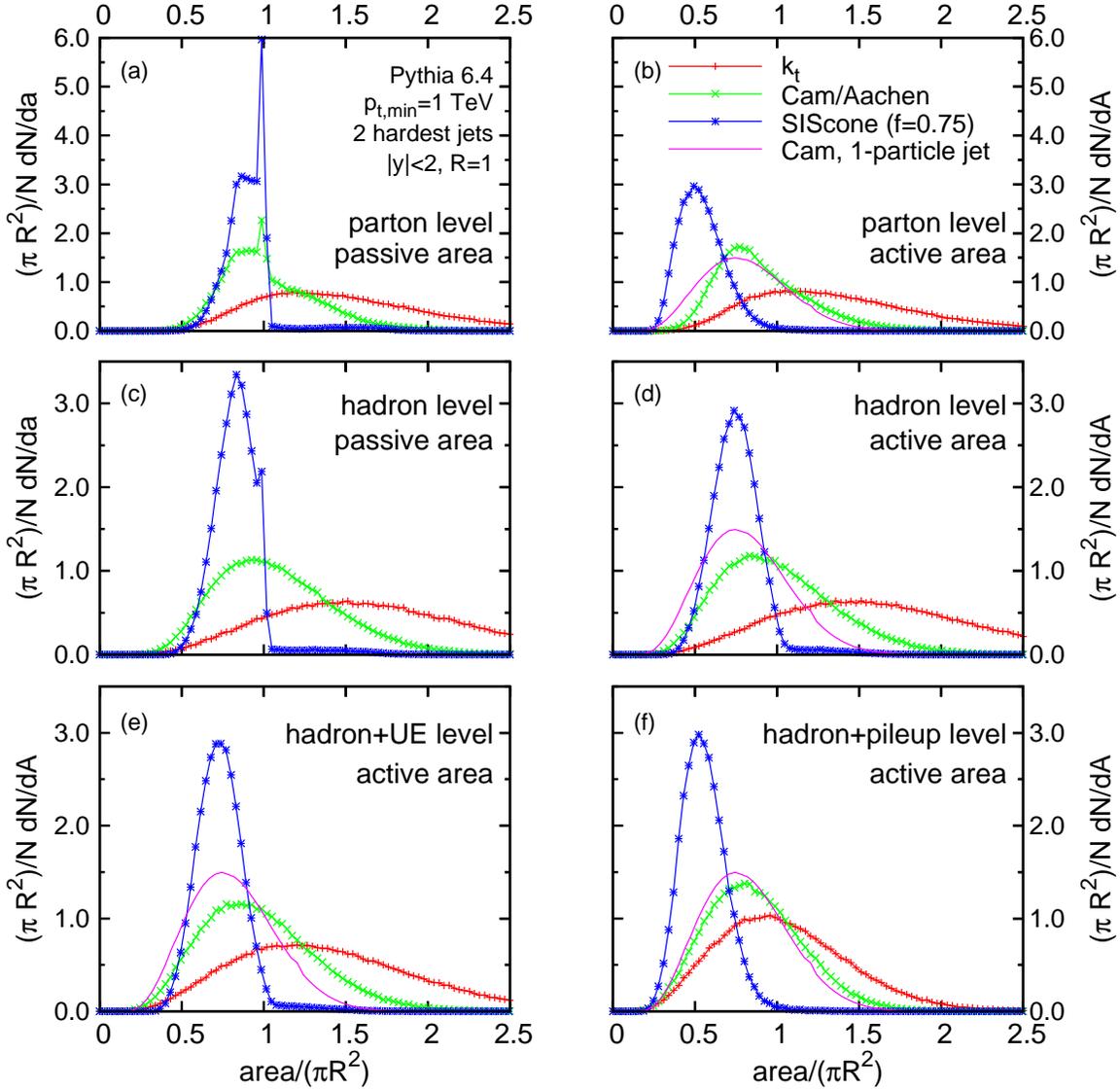}
  \caption{Distribution of active and passive areas of the two hardest
    jets in a range of simulated LHC dijet events, with a minimum
    $p_t$ of $1\TeV$ in the Pythia \cite{Pythia} event generation
    (version 6.4, default tune).
    Only jets with $|y|<2$ have been included in the histogram;
    `parton' indicates parton-level, `hadron' indicates hadron-level
    with the UE switched off, `UE' corresponds to hadron-level with UE
    switched on and in the `pileup' case the UE-level event is
    supplemented with additional minimum-bias events corresponding to
    $0.25\mb^{-1}$ per bunch crossing (about $25$ simultaneous
    interactions). For all jet algorithms we use $R=1$.
    A ghost area $\simeq 0.02$ was used throughout except for
    SISCone, where the ghost area was roughly $0.1$ in the active area
    cases.
    Note that ``area'' in these plots corresponds to
    $a_{\JA,R}(J_i)$ and $A_{\JA,R}(J_i)$ respectively for the passive
    and active areas. As such the latter has been averaged over ghost
    ensembles (see eq.~(\ref{eq:act_area})) and the dispersion is a
    consequence of the event and jet structure. 
    We have found that $5$ ghost ensembles were sufficient for $k_t$
    and Cambridge/Aachen algorithms, and $3$ for SISCone (with pileup,
    1 ensemble would actually have been enough).
    In contrast, the Cam/Aachen 1-particle jet result corresponds to
    $\pi R^2/N dN/dA(\text{1-particle-jet}|\{g_i\})$ and serves to
    illustrate how the impact of variability in the event structure in
    real events has consequences rather similar to that of the
    variability of ghost-particle ensembles in the theoretical
    arguments of the previous sections.  }
  \label{fig:area-hist-pythia}
\end{figure}

Let us start with an investigation of the distribution of the areas of
hard jets with $p_t\gtrsim 1\TeV$ in simulated LHC dijet events,
fig.~\ref{fig:area-hist-pythia}.  The area distributions are shown for
various `levels' in the Monte Carlo: just after parton showering,
after hadronisation, both with and without an UE, and finally with a
substantial pileup contribution (high-luminosity LHC, $\sim 25$ $pp$
interactions per bunch crossing). We examine both passive and active
areas.

The passive areas distributions at parton-shower level,
fig.~\ref{fig:area-hist-pythia}a, are those that are most amenable to
understanding in terms of our analytical results.  Firstly one notes
that the SISCone and Cambridge/Aachen algorithms have a clear peak at
$a=\pi R^2$. These two algorithms both have the property (cf.\
section~\ref{sec:passive-area-2particle}) that the area is not
affected by moderately collinear ($\Delta_{12} < R$ for SISCone,
$\Delta_{12} < R/2$ for Cambridge/Aachen) soft particle emission.
Thus it is possible, even in the presence of parton showering (which is
mostly collinear), for the passive area to remain $\pi R^2$.
For the cone algorithm, the other main structure is a ``shoulder''
region $0.8 \lesssim a/(\pi R^2) \lesssim 1$, which coincides nicely
with the range of values that can be induced by 2-particle
configurations (cf.\ fig.~\ref{fig:2point}). A similar shoulder exists
for the Cambridge algorithm, which however additionally has jets with
$a > \pi R^2$ --- again the range of values spanned, up to $a\simeq
1.6 \pi R^2$, is rather similar to what would be expected from the
two-particle calculations. Further broadening of these distributions
at the edges is attributable to parton-level states with more than two
particles.
In contrast the parton-level passive area distribution for the $k_t$
algorithm seems less directly related to the $2$-particle calculations
of section~\ref{sec:passive-area-2particle}. This can be understood by
observing that the $k_t$ passive area is modified even by rather
collinear emissions, and the multiple collinear emissions present in
parton showers add together to cause substantial broadening of the
area distribution.

At parton shower level, there are relatively few particles in the
event and there is no obvious boundary to the jet --- the ghosts that
we add provide a way of assigning that empty area.  It is therefore
not surprising to see significant differences between the two ways of
adding ghosts, \ie the passive and active area distributions. This is
most marked for SISCone, as is to be expected from the results of
section~\ref{sec:active_1point_cone}, which showed that for a
1-particle jet the active area is $\pi R^2/4$.  There is a trace of
this result in fig.~\ref{fig:area-hist-pythia}b, where one sees that
the SISCone distribution now extends down to $A=\pi R^2/4$. There is
however no peak structure, presumably because even a highly collinear
emission gives a slight modification of the area, cf.\
fig.~\ref{fig:active_2point_cone} (the same argument given for the
absence of a peak for the passive $k_t$ area). For the
Cambridge/Aachen and $k_t$ algorithms, there is less difference
between active and passive area distributions, again as expected.

As one moves to events with more particles, for example hadron level,
fig.~\ref{fig:area-hist-pythia}c and d, the particles themselves start
to give a clearer outline to the jets. Thus the passive and active
distributions are far more similar. This is less so for SISCone than
the others, as is to be expected, for which one still sees a trace of
the peak at $a = \pi R^2$.

For events with many particles, for example with the UE
(fig.~\ref{fig:area-hist-pythia}e) and then pileup
(fig.~\ref{fig:area-hist-pythia}f) added, the difference between
passive and active area distributions is so small that we show only
the active area result. These last two plots are the most relevant to
phenomenology. A feature of both is that the dispersion is smallest
for SISCone (but significantly different from zero) and largest for
$k_t$, precisely as one would expect from eqs.~(\ref{eq:act-fluct-seq})
and (\ref{eq:Sigmaalg1}). Another feature is how for Cambridge/Aachen
the dispersion is essentially that associated with $\Sigma(0)$, the
distribution being very similar to the 1-parton active area result,
shown as the solid line. This similarity is strongest when there is
pileup: the logarithmic enhancement that enters in
eq.~(\ref{eq:sigma2_pt}) is reduced because the pileup introduces a
large value for $Q_0 \sim 20 \GeV$. In this case even the $k_t$
algorithm starts to have a distribution of areas that resembles the
$1$-parton active area result, and for SISCone one once again sees
signs of the lower limit at $A=\pi R^2/4$, the one-particle active
area result.

\begin{figure}[p]
  \centering
  \includegraphics[height=0.9\textwidth,angle=-90]{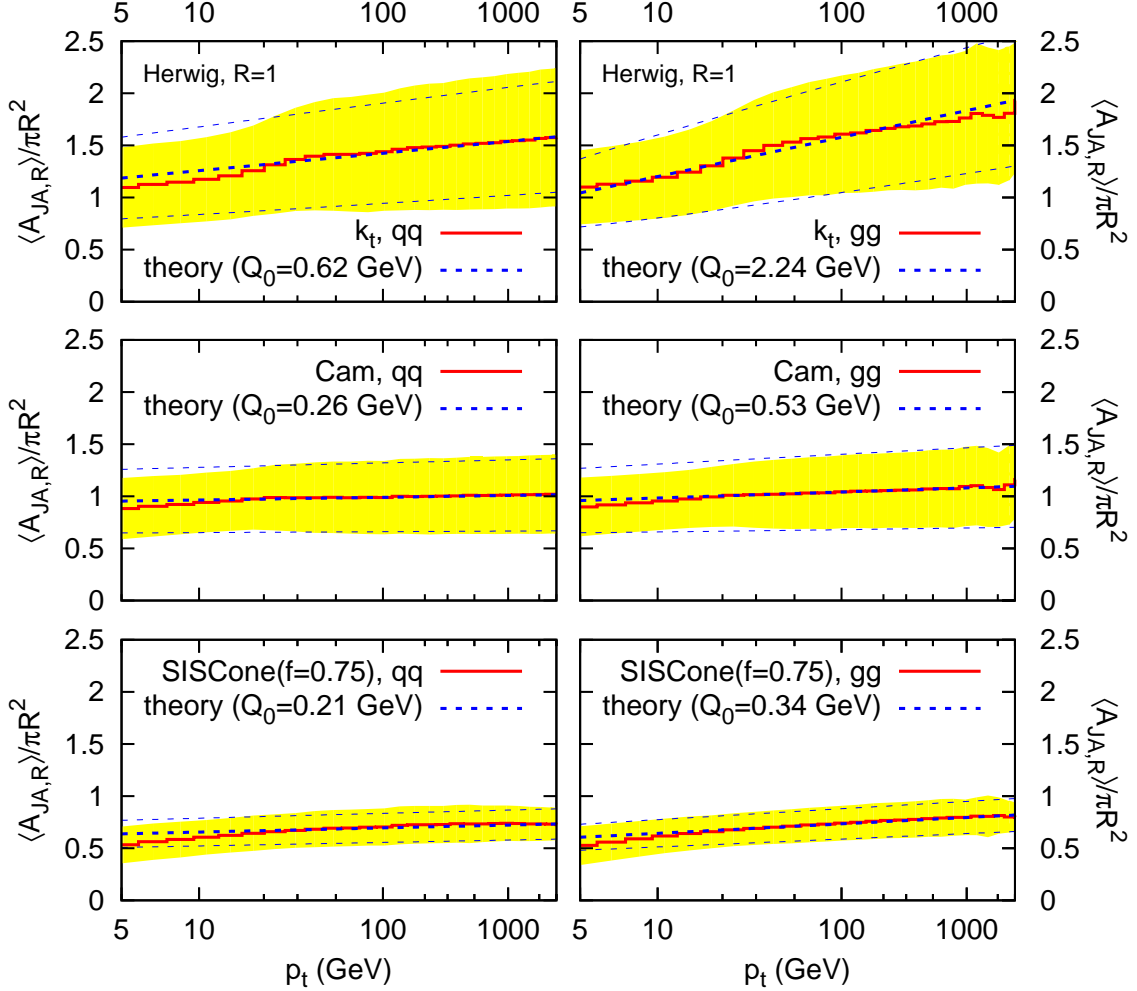}
  \caption{The mean (solid line) and standard deviation (band) of the
    active area for each of the two hardest jets in $qq\to qq$ and $gg
    \to gg$ events simulated with Herwig 6.5, as a function of the jet
    transverse momentum. The theory curves (thick dashed for mean,
    thin dashed for standard deviation) correspond to
    eqs.~(\ref{eq:Ajfr-base})--(\ref{eq:Sigmaalg1}) with a 1-loop
    coupling ($\Lambda_\mathrm{QCD} = 0.2\GeV$, $n_f=5$), and $Q_0$
    fitted (for the mean values).
    The areas have been obtained with ghosts of size $0.02$ for $k_t$
    and Cambridge/Aachen, and $0.1$ for SISCone. 
    For all algorithms, $R=1$.
    Note that the horizontal scale is uniform in $\ln \ln
    p_t/\Lambda$. Shown for $pp$ collisions with $\sqrt{s} = 14\TeV$.
  }
  \label{fig:anom_dim_herwig_gg}
\end{figure}

While the plots of fig.~\ref{fig:area-hist-pythia} provide an
illustration of many of the features that we have discussed in the
earlier sections of this paper, the fact that we have not systematically
calculated area distributions means that the discussion necessarily remains
qualitative. For quantitative checks, one should instead examine the
mean active area and its dispersion and compare them to the results of
section~\ref{sec:active}. This is done in
fig.~\ref{fig:anom_dim_herwig_gg}, separately for $qq\to qq$ and for
$gg \to gg$ scattering, as a function of the jet transverse momentum.
The horizontal scale has been chosen uniform in $\ln \ln p_t/\Lambda$
so that our predictions correspond to straight lines. 

The predictions of section~\ref{sec:active} set the slope of those
lines for each algorithm, and the agreement is reasonable in all
cases. The infrared cutoff scale $Q_0$ is not predicted, and may
differ both between quark and gluon jets and between algorithms. The
values for $Q_0$ have therefore been fitted, using the results for the
mean, and are consistent with a general non-perturbative
origin (modulo issues in SISCone discussed below). The standard
deviation (indicated by the band for the Monte Carlo simulation and
the thin dashed lines for the theory result) is then entirely
predicted, and also agrees remarkably well. Thus, overall, our simple
analytical calculations provide a surprisingly successful picture of
the mean and dispersions for various algorithms, over a range of jet
transverse momenta. This is all the more striking considering that the
calculation is based on just the first term in a series $\as^n \ln^n
p_{t}$, and is in part based on a small angle approximation.

Some remarks are due regarding the $Q_0$ values. Two clear patterns
emerge: it is largest for $k_t$, smallest for SISCone, and
systematically larger for gluon jets than for quark jets. In the case
of SISCone with quark jets, the value is uncomfortably close to the
value of $\Lambda_{QCD}=0.2\GeV$ used in the one-loop coupling. It may
be that this low value is an artefact whose origin lies in the finite
density both of actual event particles and of ghosts (the latter due
to speed limitations in SISCone): when one has a limited density of
particles and/or ghosts, the measured area may be intermediate between
the passive and ``ideal'' active areas; because SISCone has such a
large difference between passive and active areas (a factor of 4 for
the 1-particle results), the finite density effects can be
significant, and so $Q_0$ may be taking an extreme value in order to
compensate for this.

A final comment concerns the choice of event generator. Here we have
shown results from Herwig (v.~6.5). Pythia with the original
shower (the default choice in v.~6.4) is known to have difficulties
reproducing anomalous dimensions
associated with soft large angle radiation~\cite{SalamWicke,BCD},
whereas the new shower \cite{PythiaNewShower} mostly resolves these
issues \cite{BCD}.  Similar concerns are potentially relevant here
too. We actually find that both Pythia showers give results similar to
Herwig's (and ours) in all cases except the $k_t$ algorithm, for which
both Pythia showers give a slope a factor of two smaller than
expected. This suggests that an experimental measure of the $p_t$
dependence of jet areas might provide some non-trivial constraints on
parton-showering dynamics.

\subsection{Jet areas and pileup subtraction}

One of the main uses of measured jet areas is in the subtraction of
pileup \cite{AreaSubtraction}. As we have seen there are two contributions to the
modification of a jet's $p_t$ in the presence of pileup: the pileup
particles can end up in the jet, and assuming a roughly uniform
distribution of pileup (as in the case in the limit of many
simultaneous minimum-bias collisions), the resulting change in jet
$p_t$ will be proportional to the jet's area, by definition. Of course
the pileup is not exactly uniform in a given event and this must also
be accounted for. The only other modification of the jet's $p_t$ comes
from the back reaction. This gives us the following equation for the
modification of a jet's transverse momentum in the presence of pileup,
originally stated in \cite{AreaSubtraction},
\begin{equation}
  \label{eq:deltapt-for-sub}
  \Delta p_t =  A \rho \pm \sigma \sqrt{A} + \Delta p_t^{B}\,,
\end{equation}
where, as before, $\rho$ is the mean amount of transverse momentum per
unit area that has been added to the event by pileup; $\sigma$
measures the fluctuations of the pileup from point-to-point within the
event (defined as the standard deviation of the distribution of pileup
across many squares of area $1$); and $\Delta p_t^{B}$ is the net
change in transverse momentum due to back reaction. An important point
here is that $A$ is the area of the jet \emph{after} the pileup has
been added --- the fact that the area is not an IR safe quantity means
that it is not the same before and after pileup, and it is in the latter
case that it correctly measures how much pileup will have entered the jet.

The fluctuations in the jet $p_t$ due to the direct pileup
contribution can be written
\begin{equation}
  \label{eq:Delta-pt-fluct}
  \langle \Delta p_t^2 \rangle_{\PU} - \langle \Delta p_t
  \rangle_{\PU}^2 \simeq \langle \Sigma_{\JA,R}^2 \rangle \langle
  \rho\rangle^2 + \langle A_{\JA,R} \rangle\, 
  \langle \sigma^2 \rangle + \langle A_{\JA,R} \rangle^2 ( \langle
  \rho^2\rangle - \langle \rho\rangle^2)\,,
\end{equation}
where the area-related averages are those for events including the
pileup (\ie $Q_0 \sim R^2\rho$ in calculations of anomalous dimensions),
and we separately account both for point to point fluctuations of the
pileup within an event (second term), as well as fluctuations in the
overall level of pileup from event to event (third term).

\begin{figure}[t]
  \centering
  \includegraphics[height=\textwidth,angle=-90]{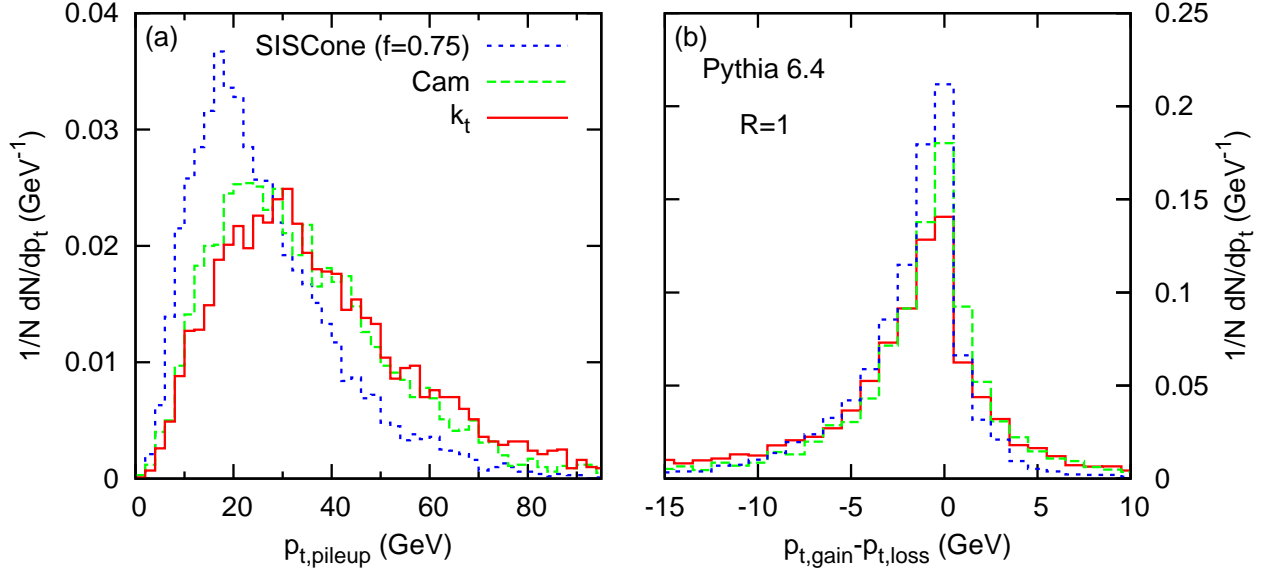}
  \caption{Both plots refer to events generated with Pythia 6.4 in
    which the two hardest jets have $p_t > 200 \GeV$ and are both
    situated at $|y|<2$. Plot (a) shows the distribution of transverse
    momentum of the pileup particles that entered each of the jets;
    plot (b) shows the back reaction on the jet, \ie the component of
    the net change in transverse momentum of the jet that is
    associated with reassignment of the non-pileup particles in the
    presence of pileup. The pileup was generated with Pythia 6.4.  
    A jet in an event with pileup is matched to the jet in the event
    without pileup with which it overlaps most, the overlap being
    defined as the transverse component of the sum of the momenta of all
    constituents that are present in both jets (\ie the method
    introduced in~\cite{CDMS}).
  }
  \label{fig:pileup_dpt_backreack}
\end{figure}

The direct pileup contribution to a sample of jets is shown in
fig.~\ref{fig:pileup_dpt_backreack}a corresponding to about $25$ $pp$
interactions per bunch crossing (high-luminosity LHC), as simulated
with Pythia tune A.
One observes that SISCone has smaller average pileup contamination and
smaller fluctuations in this contamination, which is consistent with
our analytical calculations.  More surprisingly, at least at first
sight, $k_t$ has only slightly larger and broader pileup contamination
than Cambridge/Aachen, despite its larger area. However its area is
generally larger only because of the anomalous dimension: since the
logarithm that appears in the anomalous dimension with pileup, $\log
p_t/\rho + \order{1}$, is not very large, the difference between $k_t$
and Cambridge/Aachen is essentially absent.  This is visible also in
fig.~\ref{fig:area-hist-pythia}f (the somewhat larger difference there
is a consequence of the larger jet $p_t$).

The reader may have observed that we omitted the back-reaction term in
eq.~(\ref{eq:Delta-pt-fluct}).  The tail of the distribution for
$\Delta p_t^{B} \sim p_{t,1}$ has the property $dP/d\Delta p_t^{B}
\sim \as \rho/(\Delta p_{t}^{B})^2$, so the standard deviation is
dominated by large $\Delta p_t^{B} \sim p_{t1}$, giving a contribution
to to eq.~(\ref{eq:Delta-pt-fluct}) of order $\as \rho p_{t1}$. For
large $p_{t1}$ this is parametrically bigger than the direct pileup
contributions, however it is dominated by extremely rare events (those
in which a hard particle in the jet is just near the edge of the jet),
a fraction of $\sim \as \rho/p_{t1}$ of all events. Consequently it
provides a poor estimate of the typical back-reaction.

The actual distribution for the back reaction is shown in
fig.~\ref{fig:pileup_dpt_backreack}b, and one sees that the change in
momentum is nearly always limited to a few GeV, centred close to zero.
It is slightly skewed towards loss, which is consistent with the
larger loss than gain probabilities found in
section~\ref{sec:back-reaction}.\footnote{Actually, for the $k_t$ and
  Cambridge/Aachen algorithms, one can verify that the high-$\Delta
  p_t$ tails of the distributions are consistent with the
  $dP^{(L/G)}/d\Delta p_t \simeq {\cal B}^{(L/G)} C_1\frac{2\as}{\pi}
  \frac{\rho}{(\Delta p_t)^2}$ expectation from
  section~\ref{sec:diffuse-back-reaction}.} %
Actual pileup is not perfectly uniform and for this reason the
back-reaction with SISCone is not exactly zero as would have been
expected from section \ref{sec:diffuse-back-reaction} for diffuse
radiation. Nevertheless the gain in particular is
smaller than for other algorithms, which happens to coincide with what
we would expect from fig.~\ref{fig:passive-loss-gain} for loss and
gain with pointlike pileup.

\begin{figure}
  \centering
  \includegraphics[height=\textwidth,angle=-90]{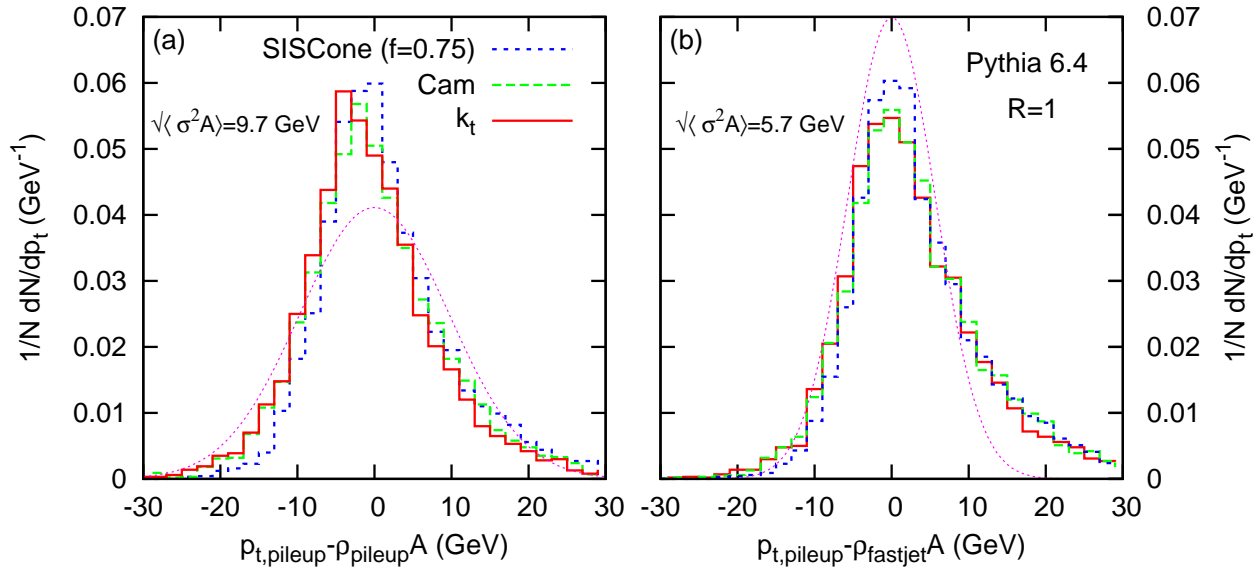}
  \caption{For the same events as in
    fig.~\ref{fig:pileup_dpt_backreack}, the transverse momentum
    contamination from pileup, minus its estimate $A \rho$ with $\rho$
    estimated event-by-event and $A$ obtained jet-by-jet (strictly
    for $A$ we have used the transverse component of the 4-vector
    area). It is compared to a Gaussian distribution with standard deviation
    $\sqrt{\langle\sigma^2  A \rangle } \simeq
    \sqrt{\langle\sigma^2 \rangle \langle A \rangle }
    $.
    In (a) $\rho$ and $\sigma$ have been estimated by splitting the
    $y$-$\phi$ cylinder into squares of area $1$ up to $|y|=3.5$,
    calculating the sum of the scalar $p_t$ of the pileup particles in
    each of them; $\rho$ and $\sigma$ are then given by the average
    and standard deviation of the contents of the squares. In (b)
    $\rho$ and $\sigma$ have been evaluated by the procedure of
    \cite{AreaSubtraction}, \ie based on median and related
    percentiles of the distribution of $p_t/A$ for all jets up to
    $|y|=3.5$, carried out on the sample of jets obtained with the
    $k_t$ algorithm with $R=0.5$.}
  \label{fig:subtraction_histogram}
\end{figure}

Let us finally come to the question of subtraction of pileup. The
equation proposed in \cite{AreaSubtraction} was 
\begin{equation}
  \label{eq:pt-correct}
  \smash{p_{tj}^{(\text{sub})}} = p_{tj} - A_j \rho \,.
\end{equation}
There we estimated $\rho$ from the event itself and effectively
subtracted both UE and pileup. Here let us estimate $\rho$ purely from
the pileup particles (with Monte Carlo simulations this is possible)
and compare the difference between the $p_t$ added by the pileup and
the $p_t$ removed by the our subtraction procedure. 
The distribution of this difference is shown in
fig.~\ref{fig:subtraction_histogram}.  The two plots correspond to two
different ways of measuring $\rho$ (either directly from the scalar
sum of $p_t$'s of pileup particles, left, or with the median-based
method of \cite{AreaSubtraction}, which is applicable also in data).
In both cases the distributions are clearly centred on zero,
indicating a successful subtraction of the pileup.

Of the fluctuation terms in eq.~(\ref{eq:Delta-pt-fluct}), only the
middle $\langle A_{\JA,R} \rangle\, \langle \sigma^2 \rangle$ piece should
remain after subtraction. This is consistent with the distributions
being much narrower than those in
fig.~\ref{fig:pileup_dpt_backreack}a. This is an important point as it
means that, in addition to subtracting the average contamination due to pileup,
we also reduce its effect on the $p_t$ fluctuations. For example, if
one wants to look at the top mass spectrum, the corresponding peak
will appear narrower (and with a better signal-to-background ratio)
after subtraction than before.
As a quantitative test, in fig.~\ref{fig:subtraction_histogram}, we
show also a Gaussian of width $\langle A_{\JA,R} \rangle\, \langle
\sigma^2 \rangle$. Here $\sigma$ has been estimated in each event
again either directly or with the methods of \cite{AreaSubtraction}.
Whereas the $\rho$ values obtained with the two methods are similar,
those for $\sigma$ differ noticeably. This may in part be due to an
occasional, hard pointlike contribution in the minimum-bias events,
causing the fluctuations measured by $\sigma^2$ to be non-Gaussian.
This would explain why the statistically more standard evaluation of
$\sigma$ shown in Fig.~\ref{fig:subtraction_histogram}a leads to
slightly too broad a distribution. In contrast the evaluation of
\cite{AreaSubtraction}, in Fig.~\ref{fig:subtraction_histogram}b, is
designed specifically to be insensitive to the pointlike component, on
the grounds that one wants a measure not influenced by the main hard
scatter. It provides a reasonable estimate of the lower half of the
distribution, but undershoots the (asymmetric) upper tail,
which one can probably ascribe to the rare pointlike component of
the minimum bias collisions.

\section{Conclusions}

The concept of a jet area is one that initially seems rather
intuitive. Yet, as we have seen in this article, there is a wealth of
physics associated with it, both in terms of how one defines it and as
concerns the interplay between a jet's internal structure and its
area. This is reflected in the range of quantities that can be studied
to characterize the behaviour of jet areas, summarised in
table~\ref{tab:all-defs}.

Our guiding principle in defining jet areas has been that they should
provide a measure of a jet's susceptibility to additional underlying
event and pileup radiation, in the limit in which these are infinitely
soft. Two opposite hypotheses for the structure of
such radiation, pointlike or diffuse, lead to two definitions of
the area, respectively passive or active, both calculated in practice
with the help of infinitely soft ``ghost'' particles. The two definitions can be
used with any infrared safe jet algorithm,
and we have applied them both
to sequential recombination algorithms ($k_t$ and Cambridge/Aachen)
and to a stable-cone with split--merge algorithm, SISCone.

The area of a jet may depend on its substructure. For the simplest
case of a jet made of a single hard particle, the passive area
coincides with one's naive expectation of $\pi R^2$ for all jet
algorithms considered here. The active area --- that related to the
more realistic, diffuse picture for UE and pileup --- instead differs
from $\pi R^2$, the most surprising result being that
for SISCone, whose active area is $\pi R^2/4$, a consequence of the
split--merge dynamics (the widely used midpoint algorithm behaves
similarly).
Thus the widespread assumption that cone-based
algorithms automatically have an area of $\pi R^2$ is, in many cases,
unjustified.

Real jets of course consist of more than a single particle. The first
level of substructure involves the addition of a soft particle to the
neighbourhood of the jet. This modifies the jet area for \emph{all} of
the algorithms considered, again including SISCone, and we have seen
that the average jet area then becomes an infrared unsafe quantity.
Nevertheless it emerges that the effects of gluon emission can
usefully be summarised in terms of an anomalous dimension, which
encodes how the jet's average area depends on its transverse momentum.
We have
calculated this to leading order and seen that it agrees remarkably
well with the $p_t$ dependence of measures of the jet areas in
hadron-level Monte Carlo simulations.

\begin{table}
  \centering
  \begin{tabular}{|l|l|p{0.58\textwidth}|}
    \hline
    quantity & discussed in & description \\
    \hline
     && \\[-10pt]
    Passive area & Section \ref{sec:passive} & single-ghost area \\
      $\bullet$ average area 
           & 
           & 
	   \\
        \hspace*{0.3cm} $a$(1PJ) 
           & Sect. \ref{sec:passive-area-2particle} 
           & \hspace*{0.3cm}passive area for a 1-particle jet\\
        \hspace*{0.3cm} $\langle a \rangle$
           & Eq.~(\ref{eq:ajfr-base}) 
           & \hspace*{0.3cm}average passive area of jets with QCD branching: \\ && 
	\hspace*{0.3cm}  $\langle a \rangle = a(\oPJ) + d\, \frac{C_1}{\pi b_0}\ln \frac{\as(Q_0)}{\as(R p_t)} + \ldots$ \\
        \hspace*{0.3cm} $d$
           & Eq.~(\ref{eq:delta-ajf-res})
           & \hspace*{0.3cm}coefficient of leading scaling violations of $\langle a \rangle$\\
      $\bullet$ area fluctuations
           & 
           & 
	   \\
        \hspace*{0.3cm} $\sigma^2(\oPJ)$
           & 
           & \hspace*{0.3cm}variance of one-particle passive area (0 by definition)\\
        \hspace*{0.3cm} $\langle\sigma^2 \rangle$
           & Eq.~(\ref{eq:passive-fluct-decomp})
           & \hspace*{0.3cm}variance of passive area of jets with QCD branching: \\ &&
	   \hspace*{0.3cm} $\langle\sigma^2 \rangle = \sigma^2(\oPJ)  + s^2 \, \frac{C_1}{\pi b_0}\ln \frac{\as(Q_0)}{\as(R p_t)} + \ldots$ \\
        \hspace*{0.3cm} $s^2$
           & Eq.~(\ref{eq:fluct_passive_coefs})
           & \hspace*{0.3cm}coefficient of leading scaling violations of $\langle\sigma^2 \rangle$
    \\[5pt]\hline 
     && \\[-10pt]
    Active area & Section \ref{sec:active} & many-ghost area (ghosts also cluster among themselves)\\
      $\bullet$ average area
           & 
           & 
	   \\
        \hspace*{0.3cm} $A$(1PJ) 
           & Eqs.~(\ref{eq:area-hard-particle}, \ref{eq:active_1point_cone})
           & \hspace*{0.3cm}active area for a 1-particle jet\\
        \hspace*{0.3cm} $\langle A \rangle$
           & Eq.~(\ref{eq:Ajfr-base})
           & \hspace*{0.3cm}average active area of jets with QCD branching: \\ && 
	   \hspace*{0.3cm} $\langle A \rangle
                = A(\oPJ)
                + D \, \frac{C_1}{\pi b_0}\ln\frac{\as(Q_0)}{\as(R p_t)} + \ldots$ \\
        \hspace*{0.3cm} $D$  
           & Eq.~(\ref{eq:Dalg1})
           & \hspace*{0.3cm}coefficient of leading scaling violations of
	   $\langle A \rangle$\\
        \hspace*{0.3cm} $A$(GJ)
           & Eq.~(\ref{eq:area-pure-ghost})
           & \hspace*{0.3cm}average active area for pure-ghost jets \\[5pt]
      $\bullet$ area fluctuations&
           & 
	   \\
        \hspace*{0.3cm} $\Sigma^2(\oPJ)$
           & Eqs.~(\ref{eq:ktcamsigmas}, \ref{eq:active_1point_cone_sigma})
           & \hspace*{0.3cm}variance of one-particle active area\\
        \hspace*{0.3cm} $\langle\Sigma^2 \rangle$
           & Eq.~(\ref{eq:act-fluct-seq})
           & \hspace*{0.3cm}variance of active area of jets with QCD branching: \\ &&
	   $\hspace*{0.3cm}\langle\Sigma^2 \rangle
                = \Sigma^2(\oPJ)
                + S^2 \,  \frac{C_1}{\pi b_0}\ln \frac{\as(Q_0)}{\as(R p_t)} + \ldots$\\
        \hspace*{0.3cm} $S^2$
           & Eq.~(\ref{eq:Sigmaalg1})
           & \hspace*{0.3cm}coefficient of leading scaling violations of
	   $\langle\Sigma^2 \rangle$\\
        \hspace*{0.3cm} $\Sigma^2$(GJ)
           & Eq.~(\ref{eq:area-pure-ghost-stddev})
           & \hspace*{0.3cm}variance of active area for pure-ghost jets
    \\[5pt]\hline 
     && \\[-10pt]
    Back reaction & Section \ref{sec:back-reaction} & action of
    finite-momentum min-bias (MB) on the clustering of the non-MB particles
    \\
      $\bullet$ pointlike MB
           & 
           & 
	   \\
        \hspace*{0.3cm} $dP^{(L,G)}/dp_{t2}$
           & Eqs.~(\ref{eq:loss-passive-master}, \ref{eq:gain-passive-master})
           & \hspace*{0.3cm}probability of jet losing (L) or gaining
           (G) $p_{t2}$ worth \\
           & & \hspace*{0.3cm}of non-MB particles\\
        \hspace*{0.3cm} $b^{(L,G)}(p_{t2}/p_{tm})$
           & Eqs.~(\ref{eq:loss-effective-area}, \ref{eq:gain-effective-area})
           & \hspace*{0.3cm}eff.\ area for loss, gain (MB
           particle has $\perp$ mom.\ $p_{tm}$) \\
        \hspace*{0.3cm} $\beta$  
           & Eq.~(\ref{eq:passive-beta})
           & \hspace*{0.3cm}coeff.~of net high-$p_{t2}$ gain$-$loss: 
           $\lim_{p_{t2}\to\infty} \frac{p_{t2}}{p_{tm}}
           b^{(G-L)}(\frac{p_{t2}}{p_{tm}}\!)\!$
           \\[5pt]
      $\bullet$ diffuse MB&
           & 
	   \\
        \hspace*{0.3cm} $dP^{(L,G)}/dp_{t2}$
           & Eq.~(\ref{eq:BR-active-loss-master})
           & \hspace*{0.3cm}probability of jet losing (L) or gaining
           (G) $p_{t2}$ worth \\
           & & \hspace*{0.3cm}of non-MB particles\\
        \hspace*{0.3cm} $B^{(L,G)}(p_{t2}/\rho)$
           & Eq.~(\ref{eq:loss-active-factorize})
           & \hspace*{0.3cm}eff.\ area for loss, gain (MB
           has $\perp$-mom.\ density $\rho$) \\
        \hspace*{0.3cm} $\cal B$  
           & Eq.~(\ref{eq:active-beta})
           & \hspace*{0.3cm}coeff.~of net high-$p_{t2}$ gain$-$loss: 
           $\lim_{p_{t2}\to\infty} \frac{p_{t2}}{\rho}
           B^{(G-L)}(\frac{p_{t2}}{\rho}\!)\!$
    \\[5pt]\hline
  \end{tabular}
  \caption{A summary of the mathematical quantities defined throughout
    this article, together with descriptions of the associated physical concepts.}
  \label{tab:all-defs}
\end{table}

\begin{table}\small
  \newcommand{\D}[1]{\comment{$#1$}}
  \centering   
  \begin{tabular}{r|cc|cc|cc|cc||cc|}
                 & $a$(1PJ) & $A$(1PJ)  & $\sigma$(1PJ) & $\Sigma$(1PJ)  & $d$         &  $D$   & $s$     & $S$      & A(GJ)  & $\Sigma$(GJ) \\\hline
   $k_t$         & $1$      & $0.81$    & $0$           & $0.28$         & $\,\,0.56$  & $0.52$ & $0.45$  & $0.41$   & $0.55$ &  $0.17$      \\ \hline
   Cam/Aachen    & $1$      & $0.81$    & $0$           & $0.26$         & $\,\,0.08$  & $0.08$ & $0.24$  & $0.19$   & $0.55$ &  $0.18$      \\ \hline
   SISCone       & $1$      & $1/4$     & $0$           & $0$            & $\!\!-0.06$ & $0.12$ & $0.09$  & $0.07$  &  ---   &  ---          \\ \hline
  \end{tabular}
  \caption{A summary of the numerical results for the main quantities
    worked out in the article and described in table~\ref{tab:all-defs}.
    All results are normalised to $\pi R^2$, and rounded
    to the second decimal figure.
    Anomalous dimensions multiply powers of $\as^n \ln^n p_t/Q_0$ that
    for typical jet transverse momenta sum to something of order $1$.
    Active-area and anomalous-dimension results hold only in the
    small-$R$ limit, though finite-$R$ corrections are small.
    Back reaction is not easily summarised by a single number, and the
    reader is therefore referred to 
    figs.~\ref{fig:passive-loss-gain}, \ref{fig:active-loss-gain}, as
    well as to fig.~\ref{fig:pileup_dpt_backreack} (right).
  }
  \label{tab:summary}
\end{table}

A summary of our main results for active and passive areas, their
fluctuations and their anomalous dimensions is given in
table~\ref{tab:summary}. As is visible also in
figure~\ref{fig:anom_dim_herwig_gg} for a broad range of $p_t$, there
is a hierarchy in the areas of the jet algorithms, $\langle
A_{\cone,R} \rangle \lesssim \langle A_{\cam,R} \rangle \lesssim
\langle A_{\kt,R} \rangle$. 
A likely consequence is that SISCone jets will be the least affected
by UE contamination, though the extent to which this holds depends on
the precise balance between pointlike and diffuse components of the
UE.
The above hierarchy might also suggest an explanation for the opposite
hierarchy in the size of hadronisation corrections, observed in Monte
Carlo studies for the different jet algorithms in \cite{CDMS}. There
the $k_t$ algorithm was seen to be least affected, which now appears
natural, since a larger jet is less likely to lose momentum through
hadronisation.

Jet areas, as well as being of interest from the point of view of
understanding contamination from UE and pileup, are also the basis of
recently developed methods for the jet-by-jet subtraction of this
contamination \cite{AreaSubtraction}. The significant fluctuations of
the area from one jet to the next mean that it is important to take
into account the area of each individual jet, rather than assume some
typical mean value. Among results here of relevance to the subtraction
procedure, we highlight the demonstration that all areas (passive,
active) are identical for highly populated events, which is important
in ensuring that the subtraction procedure is free of significant
ambiguities; we also remark on the calculation of back-reaction of
pileup on jet structure, which is found to be a small effect.

There are many avenues for potential further study of jet areas. The
calculations presented here have extracted only the leading
logarithmic part of the first non-trivial order in $\as$, and usually
we have concentrated on properties of the mean area. 
The Monte Carlo results in section~\ref{sec:real-life} also suggest
interesting structures in the distributions of areas,
%
and these merit further investigation.
Another question for future work is that of the transition
between passive and active areas if one considers a continuous
transformation of the ghosts from pointlike to diffuse. 
Since real UE and pileup contamination is neither fully
pointlike, nor fully diffuse, this question is of
particular relevance for the stable-cone type algorithms, for which
there is a large difference between the passive and active areas.

Finally, an understanding of the behaviour of jet areas can play a key
role in the choice of parameters for jet algorithms, as well as in the
design of new algorithms. One example of this concerns SISCone, for
which we saw in section~\ref{sec:active_1point_cone} that a
split--merge overlap threshold $f \simeq 0.5$ can lead to the
formation of ``monster jets,'' whereas a choice of $f \simeq 0.75$,
eliminates the problem, suggesting that the latter is a more
appropriate default value.
Another example is
given in a companion paper \cite{antikt}, where we show how to build
an infrared and collinear safe
jet algorithm, dubbed anti-$k_t$, whose passive and active areas for
single-particle jets are both $\pi R^2$, and for which the area
anomalous dimension is zero to all orders in $\alpha_s$.  These are
precisely the properties that one imagines for an ideal cone algorithm
--- but it is only once one has the tools for a quantitative
discussion of jet areas that one may establish whether a given
algorithm actually has these properties.

\subsection*{Acknowledgements}

GS thanks the LPTHE, GPS thanks Brookhaven National Laboratory and
both thank New York University for hospitality while part of this work
was carried out.
This manuscript has been authored under Contract No. DE-AC02-98CH10886
with the U.S. Department of Energy and supported by the French ANR
under contract ANR-05-JCJC-0046-01.

\appendix
\section{Definitions of the three jet algorithms}\label{sec:app_defs}

Throughout this paper, we have considered three jet algorithms: $k_t$
\cite{kt}, Cambridge/Aachen \cite{cam} and SISCone \cite{siscone}.

The first and the second of these are sequential recombination
algorithms. They introduce a distance $d_{ij}$ between each pair of
particles and a beam distance $d_{iB}$ for each particle. At each
step, the smallest distance is computed. If it involves a pair of
particles, they are recombined using a $E$-scheme sum of four-momenta
(\ie direct addition of the four-momenta), otherwise the particle is
clustered with the beam and called a `jet'. The definition of the
distance for the $k_t$ algorithm is
\begin{equation}
  \label{eq:dij-kt}
  d_{ij} = \text{min}(k_{ti}^2,k_{tj}^2)
  \frac{\Delta y_{ij}^2 + \Delta \phi_{ij}^2}{R^2}\,,\qquad
  d_{iB} = k_{ti}^2\,,
\end{equation}
where $\Delta y_{ij} = y_i - y_j$ and $\Delta \phi_{ij} = \phi_i -
\phi_j$. For the Cambridge/Aachen algorithm, the distance measures are
\begin{equation}
  \label{eq:dij-cam}
  d_{ij} = \frac{\Delta y_{ij}^2 + \Delta \phi_{ij}^2}{R^2}\,,\qquad
  d_{iB} = 1\,.
\end{equation}

The SISCone jet algorithm is an infrared- and collinear-safe
implementation of a modern cone algorithm. It first finds all stable
cones of radius $R$, a stable cone being a circle in the $(y,\phi)$
plane such that the $E$-scheme sum of the momenta inside the cone
points in the same direction as the centre of the cone. It then runs a
Tevatron run~II type~\cite{Blazey} split--merge procedure to deal with
overlapping stable cones. The stable cones are ordered in $\tilde
p_t$, the scalar sum of the $p_t$ of a cone's constituents, to produce
the initial list of protojets. One takes the hardest protojet and finds the
next hardest one that overlaps with it (its scaled transverse momentum
being $\tilde p_{t,j}$).  If the shared $\tilde p_t$ is larger than
$f\tilde p_{t,j}$ ($f$ is the split--merge overlap threshold parameter
for the algorithm), they are merged, otherwise, the common particles
are attributed to the protojet with the closer centre. If no overlaps
are found, the protojet is added to the list of jets and removed from the
list of protojets.
This is repeated until no protojets remain.

\section{Transition from one-particle jet to soft jet.} \label{sec:transition}

Having observed the different properties of the area of pure ghost
jets and jets containing a hard particle, one may wonder what happens
to the area of a jet containing a ``trigger'' particle whose
transverse momentum $p_t$ is progressively reduced, until it becomes
much smaller (ultrasoft) than the momentum scale of a soft background, $\pi R^2
\rho$ where, in the notation of section \ref{sec:real-life}, $\rho$ is
the transverse momentum density per unit area of the soft
background.\footnote{Imagine the trigger particle being a $B$ meson,
  which you could tag through a secondary vertex --- you could then
  recognise its presence in any jet, regardless of its transverse
  momentum relative to the background.}

\begin{figure}
  \centering
  \includegraphics[width=0.5\textwidth]{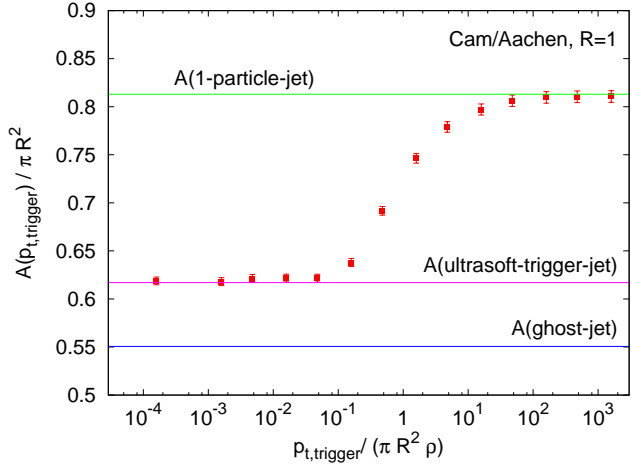}
  \caption{Area of the jet containing a trigger particle when it is
    immersed in a bath of soft particles with transverse momentum
    momentum density $\rho$; shown as a function of the trigger
    particle transverse momentum.}
  \label{fig:soft-hard}
\end{figure}

Figure \ref{fig:soft-hard} shows, as one expects, that there are two
asymptotic regions. At large $p_t$, the area tends to that of a jet
with an infinitely hard particle, eq.~(\ref{eq:area-hard-particle}).
At small $p_t$, it tends not to the pure ghost jet area,
eq.~(\ref{eq:area-pure-ghost}), as one might naively expect, but
rather to a value which can be predicted as
\begin{equation}
  \label{eq:area-ghost-biased}
  A(\text{ultrasoft-trigger-jet}) = { { \int dA \, A^2\, dN/dA}\over {\int
      dA\, A\, dN/dA }} \, ,
\end{equation}
where $dN/dA$ is the distribution of the number of pure ghost jets
with a given area, and corresponds to the solid curves depicted in
figure~\ref{fig:ghost-areas}. This equation can be understood in the
following way: when the momentum of the trigger particle becomes
negligible compared to that of the soft background, it does not
influence the size of its own jet. However, the likelihood of the
trigger particle being found in a soft-background jet is proportional
to that soft-background jet's area and one gets an extra factor of $A$
in the integrand of the numerator of eq.~(\ref{eq:area-ghost-biased}).
One is able to use the pure-ghost-jet area distribution in
eq.~(\ref{eq:area-ghost-biased}), because it coincides with that of
soft background jets.

Figure~\ref{fig:soft-hard} helps to illustrate the point that a
diffuse background of particles such as pileup (represented here by
the ghosts) provides an effective cutoff scale, below which a particle
will not have any effect on the jet's area. Numerically, the effective
cutoff does indeed coincide with the soft-transverse momentum scale
$\pi R^2 \rho$.

\section{Fluctuations of the active area}\label{sec:app_fluct}

In this appendix, we derive the results
eqs.~(\ref{eq:sigma2_pt})--(\ref{eq:sigma2_pt_end}) for the 
fluctuation coefficient $S^2_{\JA,R}$ in the case of active areas.
This is slightly more technical than for passive areas as we also have
to deal with averages over ghosts distributions. Let us briefly recall
our notation: $\avg{\cdots}$ represents an average over perturbative
emission, while $\avg{\cdots}_g$ is an average over ghosts ensembles.
When a quantity is evaluated for a specific ghost ensemble $\{g_i\}$,
we will explicitly state so with notation of the form
$A_{\JA,R}(\cdots\!\gghosts)$. From section~\ref{sec:active} we have
the implicit notation that a quantity specified without any mention of
ghosts, such as $A_{\JA,R}(0)$ is already averaged over ghost
ensembles.

At order $\alpha_s$, the mean active area can then be written (using
$A_{\JA,R}(\Delta=0)=A_{\JA,R}(\text{one particle})$)
\begin{eqnarray}
  \label{eq:def_avg}
\avg{A_{\JA,R}} &\equiv& \avg{\avg{A_{\JA,R}(\cdots\!\gghosts)}}_g
\\
 & \simeq & \avg{A_{\JA,R}(0\gghosts)
     + \int dP\, [A_{\JA,R}(\Delta\gghosts)-A_{\JA,R}(0\gghosts)]}_{\!g}\\
 & \simeq & A_{\JA,R}(0) + D_{\JA,R} \frac{C_1}{\pi b_0}
       \log\left(\frac{\alpha_s(Q_0)}{\alpha_s(Rp_{t,1})}\right),
\end{eqnarray}
with
\begin{equation}
D_{\JA,R} = \int \frac{d\Delta}{\Delta}\,
          (\avg{A_{\JA,R}(\Delta\gghosts)}_g-\avg{A_{\JA,R}(0\gghosts)}_g).
\end{equation}
Here, we have taken into account both the corrections due to the
radiation of a soft particle and the average over the distribution of
the ghosts. Note that we have used the shorthand $dP$ to represent eq.~(\ref{eq:softcoll}),
$dP/(dp_{t2} d\Delta)$, times $dp_{t2} d\Delta$, and that ``$\cdots$'' in
eq.~\eqref{eq:def_avg} represents all possible perturbative states.

For the corresponding fluctuations, the derivation goes along the same
line
\begin{eqnarray}
\avg{\Sigma^2_{\JA,R}}
 & = & \avg{\avg{A^2_{\JA,R}(\cdots\!\gghosts)}}_g - \avg{A_{\JA,R}}^2 \\
 & = & \avg{A^2_{\JA,R}(0\gghosts)+\int dP\,
       [A^2_{\JA,R}(\Delta\gghosts)-A^2_{\JA,R}(0\gghosts)]}_{\!g} \nonumber \\
 & & - \avg{A_{\JA,R}(0\gghosts)+\int dP\,[A_{\JA,R}(\Delta\gghosts)-A_{\JA,R}(0\gghosts)]}_{\!g}^{\!2}.
\end{eqnarray}
Neglecting the terms proportional to $\alpha_s^2$, we can write
$\avg{\Sigma^2_{\JA,R}} = \Sigma^2_{\JA,R}(0) + \avg{\Delta\Sigma^2_{\JA,R}}$,
where
\begin{equation}
 \Sigma^2_{\JA,R}(\Delta) = \avg{A^2_{\JA,R}(\Delta \gghosts)}_g -
 \avg{A_{\JA,R}(\Delta \gghosts)}_g^2\,,
\end{equation}
which for $\Delta=0$ is the leading order result. We can also write
\begin{equation}
\avg{\Delta\Sigma^2_{\JA,R}} = S^2_{\JA,R} \frac{C_1}{\pi b_0}
\log\left(\frac{\alpha_s(Q_0)}{\alpha_s(Rp_{t,1})}\right).
\end{equation}
Using straightforward algebra one obtains
\begin{eqnarray}
S^2_{\JA,R}
  & = & \int \frac{d\Delta}{\Delta}\,\left\{
        \avg{[A^2_{\JA,R}(\Delta\gghosts)-A^2_{\JA,R}(0\gghosts)]}_g 
        \right.\nonumber \\
        & & \qquad\qquad\qquad\left.
      - 2\avg{A_{\JA,R}(0\gghosts)}_g
      \avg{[A_{\JA,R}(\Delta\gghosts)-A_{\JA,R}(0\gghosts)]}_g 
      \right\}\\
  & = & \int \frac{d\Delta}{\Delta}\,
        \avg{[A^2_{\JA,R}(\Delta\gghosts)-A^2_{\JA,R}(0\gghosts)]}_g
      - 2\avg{A_{\JA,R}(0\gghosts)}_g D_{\JA,R} \\
  & = & \int \frac{d\Delta}{\Delta}\,\left[
        \avg{A_{\JA,R}(\Delta\gghosts)-A_{\JA,R}(0\gghosts)}_g^2
      + \Sigma^2_{\JA,R}(\Delta)-\Sigma^2_{\JA,R}(0)\right].
\end{eqnarray}
The second equality is a direct rewriting of the first. One gets to
the last line by rearranging the different terms of the first one.
The last two lines correspond exactly to eqs.~(\ref{eq:sigma2_pt_end})
and (\ref{eq:sigma2_pt_mid}) of
section~\ref{sec:area-scal-viol-active}.


\end{document}